\documentclass[12pt,preprint]{aastex}

\newcommand{\fxunits}{\mbox{ergs cm$^{-2}$ s$^{-1}$}}
\newcommand{\lxunits}{\mbox{ergs s$^{-1}$} }

\newcommand{\totclusters}{34~}  
\newcommand{\numclusters}{22~}  
\newcommand{\lowclusters}{13~}   
\newcommand{\nonclusters}{3~}   
\newcommand{\extsources}{60~}   

\def\lsim{\lower.5ex\hbox{$\; \buildrel < \over \sim \;$}}
\def\gsim{\lower.5ex\hbox{$\; \buildrel > \over \sim \;$}}

\begin{document}

\title{The WARPS Survey: VI. Galaxy Cluster and Source Identifications
from Phase I}

\author{Eric S. Perlman\altaffilmark{1,7,9,11,12,13,14,15}, Donald J.
  Horner\altaffilmark{2,8,11,12}, Laurence R. Jones\altaffilmark{3,11,13,14,15}, 
  Caleb A. Scharf\altaffilmark{1,10,12}, Harald Ebeling\altaffilmark{4,13,14,16}, 
  Gary Wegner\altaffilmark{5,17}, and Matthew Malkan\altaffilmark{6,15}}

\altaffiltext{1}{Space Telescope Science Institute, 3700 San Martin
  Drive, Baltimore, MD 21218, USA}

\altaffiltext{2}{Laboratory for High Energy Astrophysics, Code 662,
  Goddard Space Flight Center, Greenbelt, MD 20771, USA}

\altaffiltext{3}{School of Physics \& Astronomy, University of
  Birmingham, Birmingham B15 2TT, UK}

\altaffiltext{4}{Institute for Astronomy, 2680 Woodlawn Drive,
  Honolulu, HI 96822, USA}

\altaffiltext{5}{Department of Physics and Astronomy, Dartmouth
  College, 6127 Wilder Laboratory, Hanover, NH 03755, USA}

\altaffiltext{6}{Department of Physics and Astronomy, University of
  California, Los Angeles, CA 90024, USA}

\altaffiltext{7}{Department of Physics and Astronomy, Johns Hopkins
  University, 3400 North Charles Street, Baltimore, MD 21218, USA}

\altaffiltext{8}{Department of Astronomy, University of Maryland,
  College Park, MD 20742, USA}

\altaffiltext{9}{Department of Physics, University of Maryland -
  Baltimore County, 1000 Hilltop Circle, Baltimore, MD 21250, USA}

\altaffiltext{10}{Columbia Astrophysics Laboratory, Columbia
  University, Mail Code 5247, 550 West 120th Street, New York, NY 10027, USA}

\altaffiltext{11}{Guest Observer at Kitt Peak National Observatory. KPNO/NOAO
is operated by the Association of Universities for  Research in Astronomy
(AURA), Inc. under cooperative agreement with the National Science Foundation}

\altaffiltext{12}{Guest Observer at Cerro Tololo Inter-American Observatory. 
CTIO is operated by the Association of University for Research in  Astronomy
(AURA), Inc. under cooperative agreement with the National Science Foundation}

\altaffiltext{13}{Guest Observer at the Canada-France-Hawaii Telescope.  CFHT
is operated by the National Research Council of Canada, the Centre National de
la Recherche Scientifique de France, and the  University of Hawaii}

\altaffiltext{14}{Guest Observer at the W. M. Keck Observatory. The W.M. Keck
Observatory is operated as a scientific partnership among the California
Institute of Technology, the University of California and the National
Aeronautics and Space Administration.}

\altaffiltext{15}{Guest Observer at Lick Observatories.  Lick Observatory is 
operated by the University of California.}

\altaffiltext{16}{Observer at the University of Hawaii 2.2m telescope on Mauna
Kea.}

\altaffiltext{17}{Observer at the Michigan-Dartmouth-MIT Observatory.}

\begin{abstract}
  
  We present in catalog form the optical identifications for objects from the
  first phase of the Wide Angle ROSAT Pointed Survey (WARPS).  WARPS is a
  serendipitous survey of relatively deep, pointed ROSAT observations for
  clusters of galaxies.  The X-ray source detection algorithm used by WARPS is
  Voronoi Tessellation and Percolation (VTP), a technique which is equally
  sensitive to point sources and extended sources of low surface brightness. 
  WARPS-I is based on the central regions of 86 ROSAT PSPC fields, covering an
  area of 16.2 square degrees.  We describe here the X-ray source screening
  and optical identification process for WARPS-I, which yielded 34 clusters at
  $0.06<z<0.75$. Twenty-two of these clusters form a complete, statistically
  well defined sample drawn from 75 of these 86 fields, covering an area of
  14.1 square degrees, with a flux limit of $F (0.5-2.0 {\rm ~keV}) =  6.5
  \times 10^{-14} {\rm ~erg ~cm^{-2} ~s^{-1}}$.  This sample can be used to 
  study the properties and evolution of the gas, galaxy and dark matter
  content  of clusters, and to constrain cosmological parameters.  We
  compare in detail the identification process and findings of WARPS to those
  from other recently published X-ray surveys for clusters, including RDCS,
  SHARC-Bright, SHARC-south and the CfA 160 deg$^2$ survey.  
  

\end{abstract}

\section{Introduction}
\label{sec:intro}

Clusters of galaxies represent the largest gravitationally bound structures in
the universe. Their space density and evolution strongly constrain hierarchical
structure formation models, which postulate that the most massive clusters form
via mergers of less massive clusters.  The measurement of the X-ray temperature
and luminosity functions (XTF and XLF) of clusters of galaxies are powerful
discriminants between hierarchical models of structure formation (e.g.,
\citealt{nk91}), which can provide strong constraints on cosmological models
(e.g., \citealt{bahcen92,vialit96,Car97}).  In order to study the distribution
of massive clusters in the universe, as well as their characteristics, it is
crucial to compile large, relatively unbiased samples of clusters.

One of the least biased ways to select a sample of clusters with properties
directly translatable to the size of the mass aggregation, is via X-ray
surveys.  The X-ray emission in clusters originates in primordial gas and
material stripped from the galaxies in the clusters, and scales directly with
the density and temperature of the emitting gas (\citealt{JoFo84}), which
in turn are strongly correlated with the gravitating mass 
\citep{hms99}.  Optical surveys, by comparison, are strongly affected by
projection effects \citep{lu83,StRo91,Suth88,Frenk90, vH97}.  Moreover, a
number of studies have found that the optical properties of clusters (e.g.,
richness) are not very good indicators of cluster mass \citep{Frenk90}.

The Wide-Angle ROSAT Pointed Survey (WARPS; \citealt{warpsI}, hereafter Paper
I; \citealt{warpsII}, hereafter Paper II; \citealt{warpsIII}, hereafter Paper
III; \citealt{warpsIV}, hereafter Paper IV, \citealt{warpsV}, hereafter Paper
V) was designed to compile a representative, X-ray selected sample of clusters 
of galaxies.  Groups of galaxies and individual galaxies are also detectable in
WARPS, to somewhat lower redshifts than clusters (out to $z \sim 0.2-0.3$
and $z \sim 0.05$, respectively, compared to $z > 1$ for clusters).  WARPS uses
the Voronoi Tessellation and Percolation (VTP, \citealt{ebeling93,ew93})
algorithm to detect both extended and point-like X-ray sources.  Our
application of VTP to detect X-ray sources has been described in Paper I, as
has the survey calibration. An extensive optical follow-up program (described in
detail in Paper II) imaged not only extended X-ray sources (which are the most
promising cluster candidates) but also all blank field point X-ray sources, and
then took spectra of galaxies in the fields of cluster candidates.  We assume
that groups and clusters of galaxies form a continuous population, referred to
simply as ``clusters,'' and do not further distinguish between groups and
clusters of galaxies.

Early measurements of the cluster XLF, based on EXOSAT data (\citealt{Ed90})
and the {\it Einstein} Extended Medium Sensitivity Survey (EMSS,
\citealt{He92}), indicated strong negative evolution at $L(0.3-3.5{\rm ~keV})
>10^{44} {\rm ~erg ~s^{-1}}$, perhaps even at relatively low redshifts.  This
was seen as support for critical density, cold dark matter cosmologies (e.g., 
\citealt{vialit96}). However, a large number of recent results have changed the
picture substantially.  At low redshift ($z<0.3$), the results of the ROSAT
Brightest Cluster Sample (BCS, \citealt{bcsII}) show little if any evolution of
the cluster XLF.  At higher redshift, both new ROSAT surveys (RDCS,
\citealt{Ros98}; WARPS, Papers II, III; SHARC-south \citealt{Bur97};
SHARC-Bright \citealt{sharc2000}; CfA 160 deg$^2$ [also known as VMF],
\citealt{vmf98}; NEP, \citealt{mu00}) and re-analyses of EMSS data
(\citealt{Nic97}, Stocke et al., in preparation) are again consistent with no
evolution at $z<1$ (except perhaps at the highest luminosities; see in
particular \citealt{Ros98,vmfII}).  The reason for this
discrepancy is a subject of active debate, but contributing factors may include
underestimation of X-ray flux by the EMSS (Papers II, III), and
misidentification of some EMSS X-ray sources as clusters (\citealt{Nic97,
Rec99}).  Indeed, the most current analysis of the highest-redshift part of the
EMSS sample (\citealt{LG95}) is also consistent with no significant evolution
between $z\sim 0.33$ and $z\sim 0.75$ (see also Paper III; although for a
somewhat different view of the same literature see \citealt{nep}).

Here we describe the identification process and and present in catalog form
optical identifications for objects from the first phase of the WARPS project
(WARPS-I), which included data from 86 ROSAT Position Sensitive Proportional
Counter (PSPC) fields.  This sample formed the basis of Papers I and II.   The
optical identifications for galaxy clusters from WARPS-I are essentially
complete.  Papers III, IV, and V also include data from the second phase of the
WARPS project (WARPS-II), which includes nearly 4 times as many PSPC fields and
concentrates on the most distant clusters.  Those identifications are almost
complete and will be presented elsewhere (Horner et al. in preparation).

In \S~\ref{sec:methods} we describe briefly the survey methods and source
identification process, as well as cross-correlations with various astronomical
databases. In \S~\ref{sec:efficacy} we comment explicitly on the efficacy of
VTP as a method for detecting clusters of galaxies, and compare it with other
X-ray source detection methods currently in use.  In \S~\ref{sec:clusters} we
describe the statistically complete, flux limited WARPS-I sample of clusters of
galaxies, and also comment on individual clusters and interesting sources which
fall below the flux limit.  We conclude in \S~\ref{sec:prospects} by outlining
future directions for research with this and other cluster samples.

Unless otherwise stated, we use $q_{0} = 0.5$ and $H_{0} = 50$ km s$^{-1}$
Mpc$^{-1}$ when calculating distance dependent quantities. Similarly, we use
the 0.5--2.0 keV ROSAT PSPC band when quoting count rates, fluxes, and
luminosities unless otherwise noted.  The catalogs, finder charts, and other
information can be found on our WARPS WWW page at  {\tt
\url{http://lheawww.gsfc.nasa.gov/\~\ horner/warpsI/warpsI.html}}.

\section{Survey Methods}
\label{sec:methods}

The WARPS-I sample was drawn from 86 ROSAT PSPC pointings, covering a total
area of 16.2 deg$^2$.  These fields were selected at random from ROSAT
observations with exposure times greater than 8 ks and Galactic latitude $|b| >
20^\circ$.  Fields centered on bright star clusters or nearby bright galaxies
such as M31 were excluded.  This sample was used in Paper II.  However, a more
restrictive field selection has subsequently been applied to make WARPS-I
consistent with WARPS-II. This selection process excluded 11 fields, so that
the statistically complete sample is now drawn from 75 ROSAT fields (14.1
deg$^2$). Four fields (rp200510, rp400117, rp700257, and rp700302) were dropped
due to their high background level.  One field (rp600097) was discarded due to
the presence of the large optical target, NGC 600. Five other fields (rp800003,
rp800150a01, rp800401a01, rp800471n00, and rp800483n00) were dropped since the
pointing targets are galaxy clusters.  One ROSAT field originally included in
the sample, rp700305, was subsequently split in the archive into two separate
fields, rp700305a00 and rp700305a01, each with exposure time less than 8ks. 
Since it is our policy not to combine fields, we have dropped this field from
consideration.  Any clusters found in the dropped fields as part of the optical
follow-up are reported here, but they are not considered part of the
statistically complete sample. In Table~\ref{tab:fields}, we list all fields
included in WARPS-I.

\subsection{X-ray Source Detection}
\label{sub:x-ray}

In each PSPC field we utilized the annulus between $3\arcmin$ and $15\arcmin$
radius.  At the outer limit this ensures that the  point-spread function (PSF),
the size of which increases with radius, has a FWHM of  $<45\arcsec$, and also
avoids areas strongly shaded by the PSPC window support structure.  At the
inner limit, this avoids the target of each pointing.  We used only photons in
the 0.5--2.0 keV band for source detection and measurement of fluxes. This
minimizes both the size of the PSF and background relative to typical cluster
spectra.

The VTP algorithm \citep{ebeling93,ew93} was used for X-ray source detection.
VTP is a general method for detecting non-Poissonian structure in a
two-dimensional event distribution.  In the case of detected photons, VTP will
detect all regions of enhanced surface brightness. The implementation of VTP
has previously been described in Papers I and II; here we discuss only the most
important points.

The Voronoi tessellation is composed by defining each photon as the center of a
cell polygon whose sides form the perpendicular bisectors of the non-crossing
vectors joining the nearest neighbor photons. The surface brightness associated
with this cell equals the number of photons in the cell (one, unless several 
photons were registered in the same PSPC resolution element) times the inverse
of  the product of cell area and local exposure time.  To account for the
non-uniform exposure of each PSPC field, we constructed exposure maps in two
energy bands (0.5$-$0.9 and 0.9$-$2.0 keV), using an algorithm based on the
work of \citet{sno92}.  This resulted in an improvement of up to 10-15\% in the
computation of source fluxes.  VTP then searches for brightness enhancements by
comparing the cumulative surface brightness distribution with the expectation
from a Poisson distribution, and finds a global  threshold for the surface
brightness distribution in a given field at which the observed distribution
becomes inconsistent with random noise.  In the percolation step of the VTP
analysis, the algorithm  groups into sources all adjacent cells whose surface
brightness exceeds the  threshold value.  Sources consisting of very few
photons consistent with random fluctuations are discarded; for the remainder
VTP computes source  properties, such as source position, count rate, and
signal-to-noise ratio.

In using VTP, we set a lower limit in count rate of $3 \times 10^{-3}$ ct/s for
inclusion in the preliminary sample, corresponding to a flux of approximately
$3.5 \times 10^{-14} {\rm ~ergs ~cm^{-2} ~s^{-1}}$ in the 0.5--2.0 keV band.
However, our final limit in {\it total flux} ({\it i.e.}, corrected for the
component below our detection threshold) is $6.5 \times 10^{-14} {\rm ~ergs
~cm^{-2} ~s^{-1}}$ in the 0.5--2.0 keV band.  This flux limit is slightly
higher than that chosen in Paper II ($6.0 \times  10^{-14} {\rm ~ergs ~cm^{-2}
~s^{-1}}$) to be sure of completeness; it  was chosen as a result of the
simulations described in Paper I, which tested both the threshold at which we
could reliably label a source as extended as well as the typical sizes of the
flux corrections we applied for extended sources.

VTP was run five times on each field using different surface brightness
thresholds in order to distinguish real single sources from those composed of
blends of several sources (point-like or extended). This also reduces
uncertainties in the count rate of sources close to, and occasionally blended
with,  positive background fluctuations.  We then selected the optimal surface
brightness  threshold for each field by visually inspecting the VTP source
photon distribution for each of the five thresholds.  This procedure not only
allowed us to separate sources blended together at low percolation thresholds,
but also permitted us to identify and merge complex extended sources split into
several fainter sources at high percolation thresholds (cf. \S 3).
 
The list of deblended sources and respective VTP parameters were then passed to
an algorithm that estimates their true total flux.  To do this, one must make
assumptions about source morphology.    We chose to assume that sources are
either intrinsically point-like or follow a King profile with $\beta = 2/3$; 
normalization and core radius are free parameters detrmined from the VTP source
parameters. For each source, we then computed two flux estimates, $S_{\rm
King}$ and $S_{\rm point}$, under the above assumptions.  We classify an object
as extended if its extent parameter $f_{\rm ext}$, defined as the ratio $S_{\rm
King}/S_{\rm point}$, lies above 1.2, and  ``marginally extended'' if
$1.1<f_{\rm ext}<1.2$.


In Table~\ref{tab:highflux}, we list all sources found by VTP with fluxes $F_x
\geq 6.5 \times 10^{-14} {\rm ~ergs ~cm^{-2} ~s^{-1}}$. Table~\ref{tab:lowflux}
lists sources at lower fluxes and in fields that did not qualify for the
complete sample.  Optical images and spectra were not obtained for every X-ray
source, as described in \S 2.2, so for the non-cluster candidates ({\it e.g.,}
AGN, \S~\ref{sub:agn})our identification process  cannot be regarded as
complete. Only 40\% of all sources were identified either from the literature
or through follow-up observations.  The columns in Tables ~\ref{tab:highflux}
and ~\ref{tab:lowflux} are as follows:

\begin{enumerate}
\item The assigned source name.
\item The ROSAT pointing containing the source.
\item The right ascension (hours, minutes, seconds) in J2000
  coordinates (typical error is $15\arcsec$).
\item The declination (degrees, minutes, seconds) in J2000
  coordinates (typical error is $15\arcsec$). 
\item The raw VTP count rate in units of $10^{-3}$ counts per second.  
\item The corrected (i.e., total) VTP count rate in units of $10^{-3}$
  counts per second.
\item The Galactic hydrogen column density ($N_{H}$) in units of
  $10^{20}$ cm$^{-2}$ from \citet{dl90}.
\item The unabsorbed X-ray flux in units of $10^{-13}$ \fxunits\ in
  the 0.5--2.0 keV band.  For clusters of galaxies, the flux was
  calculated using the same iterative method used by \cite{bcsI}.  We
  assume a Raymond-Smith spectrum with a redshift appropriate for each
  cluster and metal abundance of 0.3 solar.  The temperature of the
  model was derived iteratively by constraining the cluster to obey
  the observed $L-T$ relation \citep{wjf97}.  Fluxes were obtained for AGN
  and unidentified sources assuming a power law
  spectrum with $F_\nu \propto \nu^{-1}$.  The flux for sources
  identified as stars was calculated assuming a Raymond-Smith model
  with a temperature of $2 \times 10^{6}$ K and no correction for
  Galactic hydrogen column density.
\item The extent parameter as discussed above.
\item The imaging status of the source.  The letter indicates the band
  of the optical image(s) obtained.
\item The status of the spectroscopic follow-up. A 'Y' indicates that
  spectra were taken for this source.
\item The identification of the source (e.g., cluster, AGN, star).
  Identifications in lower case were identified based on the
  literature (primarily using NED and SIMBAD) while upper case IDs are
  based on WARPS optical follow-up observations.  Sources listed as
  blends are briefly discussed in the notes column.
\item The redshift of the source, if known, or spectral type for stars.
\item Alternate names and other notes.  Sources listed as SHARC appear
  in the SHARC Bright Survey catalog \citep{sharc2000}.  Sources
  listed as VMF correspond to sources in the CfA 160 deg$^{2}$ survey
  of \citet{vmf98}.  Sources listed as RIXOS correspond to sources in
  the RIXOS survey catalog \citep{rixos2000}.  NVSS \citep{nvsscat}
  and FIRST (e.g., \citealt{first}) radio sources are also indicated.

\end{enumerate}

In Figure~\ref{fig:dssover}, we show overlays of the observed X-ray emission on
digitized plates from the POSS-I, UKST, or POSS-II surveys for all \extsources\
sources with $F_x \geq 6.5 \times 10^{-14} {\rm ~ergs ~cm^{-2} ~s^{-1}}$, that
are within the complete sample fields and were flagged by VTP as either
extended or marginally extended.  We have also included at the end of the
Figure, 4 additional X-ray sources which were classified as pointlike by VTP
but were selected as cluster candidates based on optical imaging 
(\S~\ref{sub:imaging}). As discussed in \S 4, one of these four sources turned
out to be a $z=0.722$ cluster. Finder charts for X-ray sources not identified
as cluster candidates are available on our WWW page.

\subsection {Optical/X-ray Selection of Cluster Candidates}
\label{sub:selection}

In order to select sources for follow-up, it was necessary to define selection
criteria for cluster candidates, as well as an integrated program of optical
imaging and spectroscopy.  This sub-section discusses the selection criteria,
while the next sub-section discusses the optical follow-up program and how it
was integrated with the selection criteria.  In Paper II, we detailed the
method by which we decided to image and take spectra of possible cluster
candidates. Here we review the essential points. 

\subsubsection {Selection Criteria}
\label{sub:criteria}

The primary motivation for the selection criteria was to ensure the
completeness of the sample of clusters of galaxies.  In order to ensure maximal
completeness, we followed up our deblending with a visual screening of every
X-ray source.  This is helpful because even after deblending, an X-ray source
can have a large $f_{ext}$ either by virtue of being truly extended or because
it is a blend of two or more unresolved sources too close to be resolved by the
ROSAT PSPC.   It is also helpful because a finite probability exists for a
truly extended source to masquerade as a point source in the VTP list, if for
example it is either very core dominated ({\it e.g.}, an extreme cooling flow)
or at very high redshifts.  Finally, as noted above, sources of low surface
brightness can be split up by deblending into multiple sources.  Each of these
effects is applicable not only to VTP but indeed to any algorithm of X-ray
source detection, or any algorithm which determines source extent, as discussed
in Paper II.  

This procedure allowed us to sort the X-ray sources into four categories - (1)
those flagged by VTP as extended which were very likely physically extended;
(2) those flagged by VTP as extended which were  very likely blends; (3) those
flagged by VTP as point-like which were possibly extended; and (4) those
flagged by VTP as point-like which were likely point-like. We then decided to
accept for our first cut list of cluster candidates {\it every} source which
might possibly be a cluster.  This included not only all sources flagged by our
source detection procedure as extended or marginally extended (i.e., categories
1 and 2 above), but also a significant number of sources flagged as point-like
(category 3 above).

Once a candidate source list was selected, we then designed an  optical
follow-up program which placed particular emphasis on both completeness and
reliability.   Included in this list were a number of sources which eventually
fell below the flux limit.  The reason for this is two-fold: first of all we
wanted to compile a complete cluster list, so all extended sources in phase I
were followed up regardless of whether they exceeded the flux limit; and
second, as our processing algorithms improved, VTP was rerun, and thus the
fluxes changed slightly.

To carry out our first screening, Automatic Plate Measuring Facility (APM; see
\citealt{Ir94}) measurements of Palomar $E$ and $O$ and UKST $R$ and $B_j$
plates were obtained at the positions of all X-ray sources with count rate $>3
\times 10^{-3} {\rm ~ct/s}$ (0.5$-$2 keV) within the survey area.  We also
obtained the Digitized Sky Survey images of the source fields, and overlaid
X-ray contours.  A  large fraction of the X-ray sources have obvious optical
counterparts which are immediately pinpointed in this step.  This also allows
one to  identify systematic pointing errors, which can reach $\sim 15 \arcsec$
and  varies in direction from dataset to dataset \citep{bp95}.  Once these
overlays were made, a source was accepted as a possible cluster candidate if it
met any one of three criteria:

\begin{itemize}

\item The X-ray source is extended or marginally extended and at least
one galaxy is visible on the digitized sky survey plate, where the galaxy (ies)
is not local ($z\lsim 0.01$) and of similar optical and X-ray size.  In this
case we not only obtained an image but immediately selected the field
for spectroscopy.

\item The X-ray source is extended or marginally extended and no
optical counterpart is visible within one magnitude of the plate
limit.

\item The X-ray source is point-like but no optical counterpart is
visible within one magnitude of the plate limit.

\end{itemize}

\begin{figure}
\plotone{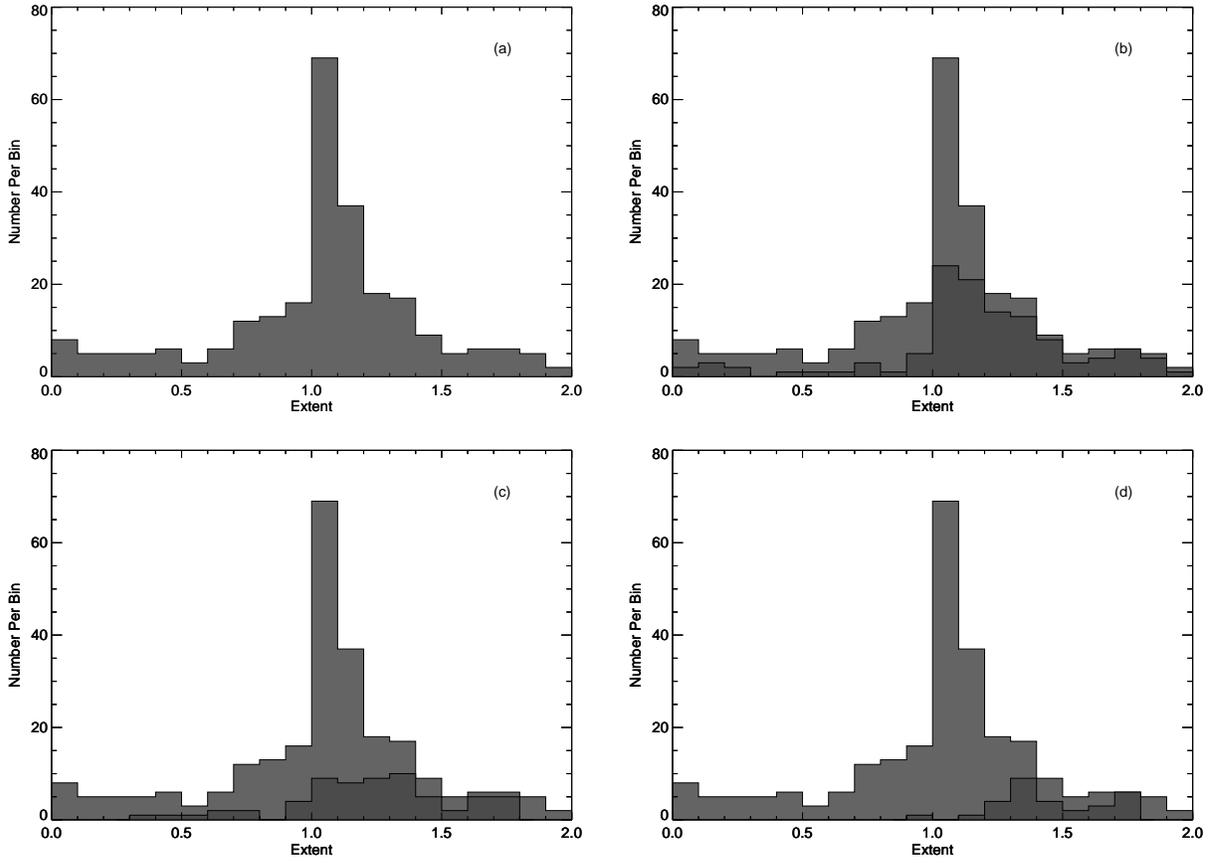}
\caption{Histogram of the extent parameter for (a) all sources, (b)
  with sources selected for imaging, (c) with sources selected for
  spectroscopy, and (d) with confirmed clusters. \label{fig:extents}}
\end{figure}

This process selected 108 sources for optical imaging.  In practice some
sources just below the flux limit were followed up because (a) the final fluxes
depended to a small degree on the unknown source spectrum, which for clusters
is based on the $L-T$ relation and thus requires knowledge of the redshift, and
(b) some source fluxes changed slightly at one point during the survey when we
reran the VTP algorithm.   Figure~\ref{fig:extents}b shows a histogram of the
extent parameters $f_{ext}$ of the cluster candidates in the complete sample
from this first screening. 

\subsubsection{Cross-correlation with other source catalogs}
\label{sub:cross}

In order to identify previously known clusters, as well as AGN, Galactic stars
and radio sources, we cross-correlated our source list with NED, SIMBAD and the
NVSS and FIRST VLA survey lists (\citealt{nvsscat}, \citealt{bwf95}).  This
produced a large number of IDs, particularly for non-cluster sources, which we
list in Table~\ref{tab:highflux} and ~\ref{tab:lowflux}.  It also allowed us to
screen out some X-ray sources we flagged as possibly extended, as more likely
AGN. The cross-correlation with the NVSS revealed that 18 WARPS clusters may
contain radio sources; the fluxes of these sources are listed in
Table~\ref{tab:radio} (see also \S~\ref{sec:clusters}). This is certainly not
unexpected; both in the present epoch and at redshifts up to $z=0.8$,
approximately 30-50\% of clusters contain radio galaxies, usually of the FR 1
type (\citealt{ol97}; \citealt{Sto99}).  In the case of optical clusters with
radio sources in the field, we have checked the X-ray/optical overlays for
X-ray enhancements and/or other evidence that some of the observed X-rays might
be emitted by an AGN.  Two such cases are noted in \S~\ref{sec:clusters}.

We have also cross-correlated our source list with those of other
ongoing ROSAT surveys, particularly those which have as their aim the
discovery of clusters of galaxies (e.g., SHARC, \citealt{sharc2000};
CfA 160 deg$^2$, \citealt{vmf98}).  The information from these
cross-correlations is given and appropriately referenced in
Table~\ref{tab:highflux}.

\subsection {Optical Follow-up Program}
\label{sub:follow-up}

Once candidates for follow-up were selected according to the procedure laid out
in \S~\ref{sub:selection}, we designed a program of optical imaging and
spectroscopy, with a further selection step in between the imaging and
spectroscopy.  The goal of the optical follow-up program was to identify all
clusters in the WARPS sample and determine their redshifts. The optical
follow-up program used many different telescopes for imaging and spectroscopy. 
Our WWW page gives details of the observations for specific sources.

\subsubsection{Optical Imaging}
\label{sub:imaging}

Images were obtained of cluster candidates, usually in the R band, which gives
a  good combination between high sensitivity, good chip cosmetics (i.e., no
fringing), and sensitivity to galaxies at all but the highest redshifts where
X-ray selected clusters are known (up to $z \sim 0.9$, where the Ca break
passes out of the R band).  These images were obtained at several different
observatories (see the WWW page for details).  Often an object was imaged more
than once, either in R or I band, if the initial images revealed a still blank
field, or evidence of faint galaxies within the X-ray flux maximum region.
Through this process, a total of 64 objects were found to coincide with an
overdensity of galaxies and thus classified as cluster candidates.

\subsubsection{Optical Spectroscopy}
\label{sub:spectroscopy}

The next step in our optical follow-up was to observe spectroscopically the 64
objects identified as cluster candidates by the method described above.  We
also observed an additional 13 sources whose IDs are listed in
Table~\ref{tab:highflux}.  In Figure~\ref{fig:extents}c, we show a histogram of the
extent parameters $f_{ext}$ of the sources selected for spectroscopy.

For each cluster candidate, we used X-ray/optical overlays to select galaxies
for spectroscopy based not only upon their brightness but also their location
relative to the X-ray centroid or peak(s).  We attempted to associate galaxy
groupings with each individual peak in the X-ray surface brightness
distribution, not only to quantify the degree of possible point-source ({\it
i.e.,} non-cluster) contamination, but also to account for superpositions of
two truly extended sources.  In obtaining spectra we typically used low
dispersion gratings, and where possible multi-slit masks were designed in order
to maximize our efficiency in identifying cluster members.

We identify an X-ray source as a cluster if (1) at least two galaxies (in the
case of two or three spectra of good signal to noise) or at least three
galaxies (in the case of four or more spectra) have very similar redshifts, and
(2) if it can be reliably determined that the X-rays were not emitted by an
AGN, star or other point source (some blends are present; see 
\S~\ref{sub:descriptions}).  As a result of this program, \totclusters\
clusters were identified, \numclusters\ of which are above our flux limit of
$6.5 \times 10^{-14} {\rm ~erg ~cm^{-2} ~s^{-1}}$.  In Table~\ref{tab:cluster},
we list information on each cluster; we will analyze the cluster population in
more depth in \S~\ref{sec:clusters}.

\subsection{AGN in the WARPS-I sample}
\label{sub:agn}

A significant number (22) of the objects we observed spectroscopically are
AGN.  These objects, along with those previously identified as AGN in other
surveys (28 found via our NED and SIMBAD cross-correlations), are listed as
such in Table~\ref{tab:highflux}.  Several of these AGN are located in fields with
significant galaxy overdensities, suggesting that those objects may reside
in a group or cluster of galaxies.  Rich environments are relatively common
among AGN, particularly radio-loud AGN (e.g., \citealt{Ell91}).  A small
fraction of the X-ray emission in objects of this sort might originate in a hot
intracluster medium (as found for several 3CRR quasars and radio galaxies by
\citealt{hw99}); however, we believe this fraction is too small to constitute a
significant source of error in our log N - log S or luminosity function. 

\section{Efficacy of VTP and Comparison with Other Surveys}
\label{sec:efficacy}

WARPS is one of several X-ray surveys for clusters of galaxies based on
archival ROSAT data.  These surveys differ operationally in two main ways: (1)
in the source detection algorithm, including the method by which sources are
classified as extended; and (2) in the optical follow-up carried out,
specifically the design of the imaging program. 

As already mentioned, in WARPS we replaced the standard sliding box detection
algorithm, which has difficulties detecting  extended sources of low surface
brightness, with the VTP algorithm (\S ~\ref{sub:x-ray}).  VTP is one of
several such algorithms currently in use for the detection of clusters in
archival X-ray data. Three of the others are based on wavelet techniques (in
particular those used by the RDCS, SHARC-Bright and CfA 160 deg$^2$ teams),
while another is based on modifying the sliding box algorithm with smoothing
and wavelet transforms (the SHARC-south survey, \citealt{Col97}). At least in
theory, both VTP and wavelet techniques should be equally sensitive to surface
brightness enhancements, regardless of their extent or morphology.  

While the basic outlines of the optical follow-up program for each survey are
similar, some important differences exist.  Chief among these are the
following. (1) WARPS is the only survey which performed optical imaging on
sources which were not flagged as possibly extended by its detection algorithm
(see \S~\ref{sub:imaging}). (2) RDCS is the only survey that has included
near-infrared imaging in its follow-up program.  (3) SHARC-Bright is the only
survey that used a completely automated 'friends-of-friends' analysis to 
screen for possibly blended X-ray sources.  Each of these could conceivably
have an impact on the completeness and content of the respective samples.

A comparison of the number counts as well as the results in overlapping fields
is useful in judging the efficacy of each procedure, and the completeness of
each sample. The efficacy of a detection algorithm in detecting extended
sources can be judged in several ways.  The first is to compare the number of
sources that were extended by eye in a given field with the number of extended
sources found by the detection algorithm.  This is what might be called the
``false negative'' detection rate.  For WARPS, such a visual inspection was
done for every field as a part of the deblending process detailed in Paper I. 
These inspections were originally done as part of an effort to distinguish
blends of two or more point sources from truly extended X-ray sources, but they
also have the side benefit of being able to find extended sources which may
have been missed by VTP.  This visual examination yielded only one more
possibly extended source, listed multiply as WARPJ1414.9+1120, WARPJ1415.0+1119
and WARPJ1415.2+1119, compared to over 30 extended sources.  It should be noted
that this source {\it was} found by running VTP at the lowest of the three
thresholds described in Paper I, but was removed by the deblending process that
paper described, and manually returned to the catalog after our by-eye
inspections.  It is also not surprising that this source is extremely faint and
in fact does not make our flux-limited sample (\S~\ref{sec:clusters}).  This
very low fraction of ``lost'' sources gives us considerable confidence in the
reliability of VTP for detecting all extended and point sources up to our
selected flux limit (for a discussion of the statistical significance of this
flux limit see Paper I).

A second way to test the efficacy of our selection method, and in particular
$f_{ext}$, described in \S~\ref{sub:x-ray} and Paper I, is to look at the
relationship between the fraction of truly extended X-ray sources (i.e.,
clusters, elliptical galaxies, etc.) and $f_{ext}$.  In
Figure~\ref{fig:extents}d we show a histogram of the fraction of sources
identified as clusters, as a function of $f_{ext}$.  As can be seen, one of the
clusters in our flux-limited sample has $f_{ext}<1.1$ (WARPJ2302.8$+$0843), and
two more likely clusters with $f_{ext} < 1$ fall below the limit
(WARPJ1537.7$+$1201 and WARPJ1501.0$-$0824, which are in high-background 
fields). These clusters were identified only because our optical follow-up
program included not only all possibly extended sources, but also sources
flagged as not extended by VTP, that might be associated with visible galaxies
or were blank fields. The cluster fraction increases markedly at higher values:
11\% of sources with $1.1<f_{ext}<1.2$ are clusters, but about 47\% of sources
with $f_{ext}>1.2$ are clusters.  A similar fraction holds at higher extents. 
The fact that it never approaches unity attests to the fact that a significant
number of X-ray blends are present.  This pattern gives us further confidence
in the efficacy of VTP as a method to detect extended sources.

It is not surprising that two of three clusters with $f_{ext}<1.1$ are among
our faintest and most distant.  Simulations by \citet{eh91} showed that
cosmological $(1+z)^{-4}$ surface brightness dimming can cause a cluster to
appear as a point source at high $z$.  Those authors predicted that ROSAT would
begin to have trouble distinguishing extended X-ray emission in clusters from 
point-like sources, at redshifts $z>0.4$.  The results of this and other X-ray
surveys for clusters of galaxies (Papers I, II;
\citealt{sharc2000,Col97,Ros98,vmf98}) have shown that this prediction
was too pessimistic, perhaps partly due to evolution of the spatial and/or
spectral distribution of the ICM in clusters with redshift (which is predicted
in hierarchical models of structure formation; see
\citealt{bahcen92,vialit96}), or the presence of substructure and less
centrally concentrated emission.  However, at particularly high redshifts this
effect should not be forgotten and may be a cause of incompleteness not only in
our survey but in all ROSAT-based surveys (see also the simulations by
\citealt{Ad00}).

The third and final way to test efficacy is to compare the overall number
counts in WARPS with those of other surveys.  If indeed all of the methods
being employed are comparably effective in finding extended X-ray sources and
hence clusters, we should find comparable log~N~-~log~S distributions. 
\citet{Ros98} have already addressed this issue regarding the comparison of
WARPS-I with RDCS: despite the fact that RDCS has found a few clusters at $z>1$
\citep{ros00}, the overall number counts are comparable.  The situation is
similar for the  160 deg$^2$ survey of \citet{vmf98}, which has a log N - log
S distribution similar to WARPS-I (\citealt{vmf98}, their Fig. 10).  Moreover,
since that survey has published its catalog we can cross-check the sources and
IDs in the fields that are in common between the two.  A total of 20 ROSAT
fields are common to WARPS and 160 deg$^2$ survey, and the sources and IDs
agree well (in particular the cluster IDs are in complete agreement in the
overlapping regimes of flux and detector area, and the fluxes also agree well,
with a mean ratio of VMF/WARPS=$0.98 \pm 0.26$).  The sources in common are
noted in Table 3. Finally comparing the number counts of WARPS-I and
SHARC-south (\citealt{Col97,Bur97}) we see that the two curves are also
comparable.

Some similar comparisons can also be made with the SHARC-Bright survey
\citep{sharc2000, Ad00}, which however have not published a 
log N - log S distribution.  
Comparing our field list with that of SHARC-Bright \citep{sharc2000} we find 60
fields in common.  WARPS finds all 6 of the clusters found by SHARC-Bright in
those  fields (down to the SHARC-Bright flux limit), as well as two not found
by  SHARC-Bright:  WARPJ1552.2+2013 and WARPJ0238.0$-$5224.  
The overall number counts also do not agree: SHARC-Bright find 37
clusters with fluxes $F_x \gsim 2.8 \times 10^{-13} {\rm ~erg ~cm^{-2}
~s^{-1}}$ (as explained in  \citealt{sharc2000}, this limit is approximate
because they perform their limit calculation in counts, rather than flux), in a
total area of 178.6 deg$^2$.  By comparison, our published log N - log S
distribution (Paper
II) predicts $57 \pm 8$ clusters within the SHARC-Bright survey - i.e., the
number counts for SHARC-Bright are too low compared with the WARPS-I, SHARC-S,
RDCS and CfA 160 deg$^2$ surveys, at the 2.5 $\sigma$ level. 

A similar difference in number counts is found when comparing the extended 
source lists of the CfA 160 deg$^2$ and SHARC-Bright surveys. Romer et al.
(2000, see their \S 7.5) note that SHARC-Bright finds only 13 of 21 CfA 160
deg$^2$ clusters that are both within the SHARC-Bright survey area and above
the SHARC-Bright flux limit.  This is consistent with the above comparison of
the WARPS-I log N - log S with the total number of galaxies from
SHARC-Bright.    Of the 8 CfA 160 deg$^2$ clusters missed by SHARC-Bright, 7
did not meet the SHARC-Bright ``filling  factor'' criterion (essentially an
automated friends of friends analysis intended to spot blends), while 1 did not
meet their extent criterion.  Thus a comparison to both our results and that of
the CfA 160  deg$^2$ results suggests either serious incompleteness, or
systematic miscalculation of cluster fluxes by SHARC-Bright.   

Romer et al. (2000) have addressed this last point by comparing the fluxes for
sources in common between SHARC-Bright and the CfA 160 deg$^2$.  They find that
SHARC-Bright find systematically higher fluxes than the 160 deg$^2$
survey, with a mean ratio SHARC-B/VMF = $1.18 \pm 0.26$, and attribute the 
difference to assuming a fixed core radius of 250 kpc (when the \citealt{vmf98} core
radius values are used the systematic difference disappears). We find a similar
average ratio SHARC-B/WARPS = $1.37 \pm 0.26$ but we are unable to test whether
this would be removed by using the SHARC-Bright pipeline with our core radii.

If the SHARC-Bright fluxes are indeed at 20-30\% variance with other,
equivalent, surveys, this would bring their overall number counts into
agreement with those of WARPS-I and the CfA 160 deg$^2$. However this still
does not explain why SHARC-Bright appears to miss $\sim 30-40\% $ of clusters
which should meet their flux limit. \citet{sharc2000} conclude that this
``illustrates how differing selection criteria produce differing cluster
samples and that  detailed simulations are required to determine a survey's
selection function.'' These were done in \citet{Ad00}, which simulated the
selection function by placing cluster sources of various X-ray morphology
randomly in  real PSPC data, and then tracking the extended source finding
rate.  But \citet{Ad00} do not test explicitly the filling factor criterion
which appears to have been responsible for SHARC-Bright missing fully 1/3 of
the clusters within their survey area.  By placing  
simulated clusters randomly in real PSPC data, \cite{Ad00} do {\it implicitly}
test the filling factor criterion, but they compile no statistics regarding
{\it e.g.}, the nearness of the simulated cluster to other sources in each
field.   The \citet{Ad00} simulations result in a selection function which
brings the SHARC-Bright X-ray luminosity function into agreement with the
results found by RDCS, WARPS and other surveys.  However, the lack of an
explicit test of the filling factor criterion makes their result more
uncertain.

Because of the exhaustive nature of the WARPS optical follow-up program
(\S~\ref{sub:follow-up}), we can make one further comment.  Unlike all the
other surveys WARPS investigated thoroughly all possible cluster sources, even
if the source was not extended according to the VTP algorithm. All other ROSAT
based surveys did not image optically any sources that were not found to be
extended by their source detection algorithm.  The similarity of the WARPS,
SHARC-south, RDCS and 160 deg$^2$ number counts, plus the fact that we only
found one cluster in our complete sample among the sources we flagged as not
extended, assures us that for our sample as well as the RDCS and CfA 160
deg$^2$ this effect is negligible overall, but may not be insignificant at the
highest redshifts, although the size of the effect depends on the details of
the algorithm used.  However, the utility of this practice becomes clear in the
comparison with SHARC-Bright (above), since the careful evaluation of
questionable sources and followup optical observations of those sources might
have enabled them to identify the missing 30-40\% of  {\it bona fide}, X-ray
bright clusters within their survey area and flux limits.


\section {Clusters of Galaxies}
\label{sec:clusters}

\subsection{The Cluster Catalog}\label{sub:catalog}

In Table~\ref{tab:cluster} we have listed all \numclusters members of
the statistically complete WARPS-I cluster sample.  Similarly, in
Table~\ref{tab:other} we list \lowclusters additional clusters and
\nonclusters possible other clusters.  Individual descriptions of all
the sources in given in \S~\ref{sub:descriptions}.  The columns in
Table~\ref{tab:cluster} and Table~\ref{tab:other} are as follows:

\begin{enumerate}

\item The assigned cluster name.
\item The right ascension (hours, minutes, seconds) in J2000
  coordinates, as extracted from the ROSAT data.
\item The declination (degrees, minutes, seconds) in J2000
  coordinates, as extracted from the ROSAT data. 
\item The extent parameter $f_{ext}$ as discussed in \S~\ref{sub:x-ray}.
\item The hydrogen column density from \citet{dl90} in
  units of $10^{20}$ cm$^{-2}$.
\item The unabsorbed source flux in units of $10^{-13}$ \fxunits\ in
  the 0.5--2.0 keV band as described in \S~\ref{sub:x-ray}.  For some
  clusters this flux may include a component from contaminating
  sources (e.g., AGN).  See \S~\ref{sub:descriptions} for details for
  each cluster. 
\item The rest-frame source luminosity in units of $10^{44}$ \lxunits\ 
  in the 0.5--2.0 keV band derived in the same manner as the flux.
  Values of $q_{0} = 0.5$ and $H_{0} = 50$ km s$^{-1}$ Mpc$^{-1}$ were
  assumed.
\item The core radius of the cluster in arcseconds.  Note that this is
  very crude as it is based on ROSAT data which in most cases barely
  resolve the cluster.
\item The redshift of the cluster.
\item The number of galaxy redshifts used to derive the cluster redshift.
\item Flags.  A ``c'' indicate that the cluster is contaminated or
  possibly contaminated by
  other sources.  In Table~\ref{tab:other}, a ``$<$'' indicates that
  the cluster flux falls below the limit for inclusion in the
  statistically complete, flux-limited sample.  An ``?'' indicates a
  possible cluster or group for which optical follow-up was not
  completed, usually because it was below the flux limit.  An ``8''
  indicates that the cluster was found in a field targeted at another
  cluster while an ``b'' indicates that this field was dropped due to
  high background.  See  the notes in
  \S~\ref{sub:descriptions} for flagged sources.
\item Alternate names or notes.  The same as for Table~\ref{tab:highflux}.

\end{enumerate}

In Figure~\ref{fig:ccdover} we show the optical images we obtained of
these clusters, with the X-ray emission contours overlaid.  These
images, as well as other information, are also available on our WWW
site.

\subsection{Radio Sources}
\label{sub:radio}

In Table~\ref{tab:radio}, we list the position, positional error,
distance from the x-ray centroid, and flux of all radio sources within
$2\arcmin$ of the X-ray centroid of each of these clusters, along with
information regarding the possible nature of the radio sources (see
also \S~\ref{sub:cross}). In each case we have checked to
see whether some fraction of the X-ray emission could come from the
radio source, and make comments to this effect in
\S~\ref{sub:descriptions}.

\subsection{The Statistically Complete WARPS-I Sample}
\label{sub:complete}

We have selected from the sources identified as clusters a statistically
complete, X-ray surface brightness limited sample of \numclusters\ clusters. 
This sample is somewhat different from the sample used to derive the
log~N~--~log~S in Paper II.  To make the WARPS-I sample consistent with that
from WARPS-II, the PSPC field selection criteria for the statistical sample
were made more restrictive (see \S~\ref{sec:methods}), and all the fields were
reprocessed by VTP.  The VTP algorithm was improved and refined between the
WARPS-I and WARPS-II processing.  This has changed the fluxes and extents
slightly, by 5-10\%, for some sources.  A few clusters used in Paper II now
fall below the flux limit (while others are now above it).  However, these
changes do not affect the conclusions of Paper II.

\begin{figure}
\plotone{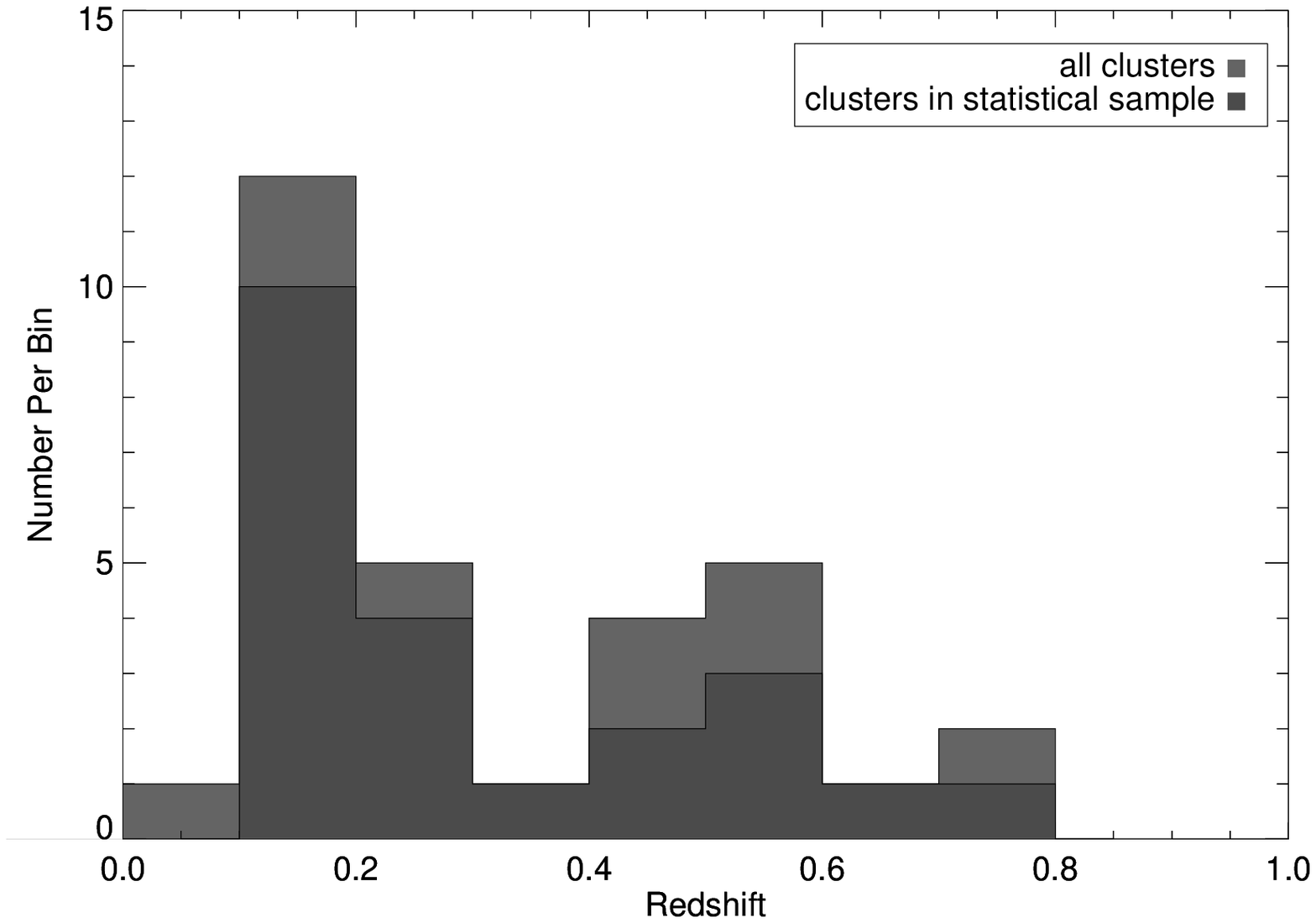}
\caption{Redshift distribution of the cluster sample.\label{fig:zdist}}
\end{figure}

Figure~\ref{fig:zdist} shows a histogram of the redshift distribution of the
sample.  As can be seen, the clusters in the WARPS-I sample range in redshift
from $z=0.06-0.75$ (see also Table~\ref{tab:cluster}).  

\subsection{Descriptions of Individual Clusters}
\label{sub:descriptions}

Here we provide detailed information about clusters in both the statistical and
supplemental samples.  See Figure~\ref{fig:ccdover} for optical CCD images with
X-ray emission contours overlaid.  In this section, SHARC-B refers to the SHARC
Bright Survey catalog \citep{sharc2000}, VMF to the CfA 160 deg$^{2}$ survey
\citep{vmf98}, and RIXOS to the RIXOS survey catalog \citep{rixos2000}.

\begin{description}
  
\item[WARPJ0022.0$+$0422] This extended X-ray source is associated
  with a cluster, GHO00190.5$+$0405 (no previous redshift), which was
  previously discovered by \citet{gho86}.  An HRI observation of this
  field failed to detect the cluster, but it is near the edge of the
  HRI field-of-view.  The northeastern X-ray component is identified
  with an AGN at $z=0.272$, unrelated to the cluster.  Aperture
  photometry on the PSPC image suggests that the quoted flux includes
  a $\approx$25\% contribution from the AGN. This cluster is in a
  ROSAT field targeted at another cluster (GHO 0020+0407 at z=0.698)
  and so is not part of the statistically complete sample.
  
\item[WARPJ0023.1$+$0421] This X-ray source originates in a poor
  cluster of galaxies at $z=0.453$.  An HRI observation of this field
  shows a marginal detection of an extended source.  All the 7
  galaxies for which spectra were obtained, including the 3 cluster
  members, show evidence of narrow emission lines. The brightest
  cluster member, offset from the X-ray peak, has \ion{O}{2} emission
  and strong H$\delta$ absorption, indicative of recent star formation
  activity.  The cluster galaxy at the X-ray peak has strong
  \ion{O}{2} and weaker H$\beta$ and \ion{O}{3} emission, indicative
  of a cooling flow or LINER-type spectrum rather than a Seyfert 2.
  This cluster is in a ROSAT field targeted at another cluster (GHO
  0020+0407 at z=0.698) and so is not part of the statistically
  complete sample.

\item[WARPJ0111.6$-$3811] The galaxies associated with this X-ray
  source appear to fall into two regions: a northern cluster at $z =
  0.121 ~(n_z = 3)$ and a south western one at $z=0.132 ~(n_z = 4)$.
  The x-ray emission originates predominantly from the northern
  cluster (the southern cluster clearly falls well below our flux
  limit).  The VTP flux estimate excludes the southern cluster and the
  bright point-like X-ray source to the southeast, which is identified
  with a QSO at $z=0.894$.
  
\item[WARPJ0144.5$+$0212] This clearly extended X-ray source
  corresponds with the position of a poor cluster of galaxies at
  $z=0.165$, revealed by the optical image (Figure~\ref{fig:ccdover}).
  The extended nature of the X-ray emission and the position of the
  BCG coincident with the X-ray peak makes it clear that the majority
  of the X-ray emission comes from the cluster, but the presence of an
  80 mJy radio source (Table~\ref{tab:radio}) makes some caution
  advisable.  The radio source corresponds to one of the fainter
  galaxies very near the BCG, which may be a background radio galaxy
  or other AGN.  However, we see no evidence for X-ray point-like
  emission at this position so the possible contribution from the
  radio source is small.  A second, much fainter, radio source is
  located in this cluster's outskirts. The quoted flux includes both
  northern and southern X-ray components.
  
\item[WARPJ0206.3$+$1511] Our redshift is from two galaxies but agrees
  with the redshift in the RIXOS catalog and the estimate in the VMF
  catalog. The quoted X-ray flux may include a small contribution from
  a bright star to the east of the cluster.
  
\item[WARPJ0210.2$-$3932] The cluster was discovered during the
  reprocessing and corresponds to VMF 24.  VMF estimate a photometric
  redshift of $z=0.19$.  Since it is well below our flux limit, no
  spectroscopic follow-up was done.

\item[WARPJ0210.4$-$3929] Our redshift for this cluster ($z=0.273$) is 
  tentative, being based on low signal to noise observations at the CTIO 1.5m
  and at the AAT in service time.  In both instances, spectra were taken of
  the  two galaxies near the X-ray centroid.  Those spectra show a continuum
  break at about 5000 \AA~ in both galaxies and probable CaII at 4994/5059 \AA~
  and H$\beta$ at 6203 \AA~ in the brighter galaxy only.  This cluster is also
  in the VMF catalog, who estimate a photometric redshift of $z=0.27$ with a
  range of 0.20-0.30.  Further spectroscopic observations are  required to
  confirm the redshift of this cluster.

  
\item[WARPJ0216.5$-$1747] This clearly extended X-ray source corresponds to the
  position of the optical cluster shown in Figure~\ref{fig:ccdover}.  Also in
  the X-ray error circle (and near X-ray peaks) are 2 M stars.  A 7.2 mJy radio
  source (Table~\ref{tab:radio}) is about $100\arcsec$ north of the X-ray
  centroid (and well outside the X-ray emission contours) which is associated
  with a foreground 14th magnitude galaxy.  The X-ray image shows no sign of
  point source contamination, and we estimate that $<$30\% of the PSPC flux is
  from the M stars, consistent with the $\alpha_{ox}$ of M stars in
  \citet{Sto91}.  

\item[WARPJ0228.1$-$1005] This clearly extended X-ray source is
  associated with a cluster which is dominated by two bright galaxies
  that overlap on the sky.  We only have a redshift ($z=0.149$) for one
  of them.  The spectrum of the other galaxy contains only very weak
  absorption lines, and may possibly be a BL Lac object. It is a also
  faint radio source (7 mJy, Table~\ref{tab:radio}).  The HRI image
  confirms the cluster to be an extended X-ray source and also shows
  an unrelated point X-ray source $\approx 1\arcmin$ to the northwest
  of the cluster.  The quoted PSPC flux includes a $\approx$10\%
  contribution from this source.
  
\item[WARPJ0234.2$-$0356] This distant ($z=0.447$) cluster falls just
  below the flux limit of our sample at $6.4\times 10^{-14}$ \fxunits.
  The X-ray contours and the galaxy velocities (although from only 5
  cluster members) suggest that the cluster has two components.  The
  BCG of the northern component has narrow \ion{O}{2}, \ion{O}{3} and
  H$\alpha$ emission lines, as well as strong H$\delta$ absorption,
  perhaps indicating recent star-formation activity. It may be
  interacting with a nearby galaxy.


\item[WARPJ0238.0$-$5224] This massively extended X-ray source corresponds with
  the cluster Abell 3038.  This source is also in the VMF and SHARC-B catalogs,
  but only in one of two fields within which the cluster is actually found (see
  \S ~\ref{sec:efficacy}.  We used the redshift from \citep{Nes90},
  which has been confirmed by SHARC. An HRI image confirms the lack of
  point-source contamination.

\item[WARPJ0250.0$+$1908] We have only poor quality optical images of this
  source and two redshifts.  However, this source is also in the SHARC-B
  catalog, who list a similar redshift ($z=0.12$).  The northern X-ray source,
  visible on the contours, is excluded from the quoted X-ray flux.

\item[WARPJ0255.3$+$0004] This marginally extended source was imaged with the
  WIYN 3.5m revealing a possible distant group or poor cluster, but as it was
  below the flux limit, no further follow-up was done.

\item[WARPJ0348.1$-$1201] This extended X-ray source emanates from the
  unrelaxed $z=0.4876$ cluster shown in the optical image. The brightest
  galaxies in this cluster fall along a southeast to northwest line.  One of 
  these is a 34 mJy radio source (Table~\ref{tab:radio}), but no X-ray
  point-like emission is evident near the position of this radio source.
  
\item[WARPJ0951.7$-$0128] This clearly extended X-ray source is
  associated with a $z=0.567$ cluster of galaxies.  Our redshift
  agrees very well with the photometric estimate published in the VMF
  catalog. Our cross-correlations with the NVSS and FIRST revealed a
  6.2 mJy radio source near the X-ray centroid
  (Table~\ref{tab:radio}).  The optical counterpart to this radio
  source is a 20th magnitude galaxy, likely a cluster member.  We see
  some evidence for unresolved X-ray emission near this position, so
  some of the X-ray emission may originate from the radio source.  We
  have only one firm redshift for this cluster and one other probable
  similar redshift.

\item[WARPJ1113.0$-$2615] The galaxies close to the brightest galaxy in this
  distant ($z=0.725$) cluster show morphological evidence for interactions, and
  perhaps a gravitationally lensed arc.  The X-ray source is clearly extended 
  but low surface brightness and is very close to the flux limit for our
  statistically complete sample.  Based on visual inspection of our deblending
  results, we have used the count rate at threshold 3 (rather than 4) for this
  object.

%
  
\item[WARPJ1406.2$+$2830] We obtained only a poor image and one low quality
  redshift of $z=0.552$ for the BCG of this source.  This source is also listed
  in the CRSS survey \citep{bwe97}, but as a galaxy only;  the CRSS give a
  redshift of 0.546, also based on a low quality spectrum.   VMF list this 
  source with the CRSS redshift, but also identify the source as a cluster.
  We observed this cluster with the HRI in 1996,
  with exposure time 19900 seconds, but the HRI failed to detect extended
  emission from the cluster or any point source contamination. Given the lower
  sensitivity of the HRI compared to the PSPC to low surface brightness X-ray
  emission, this result supports supports the assertion that this X-ray source
  is indeed extended.

\item[WARPJ1406.9+2834] An HRI image shows an extended source at the
  position of the cluster plus a point X-ray source $90\arcsec$ east
  of the cluster center.  The flux quoted includes the flux from the
  point source, but analysis of the HRI image indicates that the
  point-source flux is at most 25\% of the cluster flux.  This source
  is also in the VMF and SHARC catalogs.

\item[WARPJ1407.6$+$3415] This extended X-ray source falls somewhat below our
  statistically complete sample's flux limit.  We have one low quality redshift
  at $z=0.577$ and another at $z=0.3793$, which contains narrow emission lines
  (\ion{O}{2} equivalent width is 20{\AA}).  This may be a cluster contaminated
  by a foreground star-forming galaxy or LINER.  No further follow-up was done.
  
\item[WARPJ1415.1$+$3612] This distant cluster contains a 5.4 mJy
  radio source near the center of the X-ray contours possibly
  associated with the cluster BCG.  The X-ray morphology indicates
  possible point-source contamination.  We have obtained optical spectroscopy
  of this cluster with Keck but the data are noisy and we have been unable to
  extract a redshift.
  
\item[WARPJ1415.3$+$2308] This poor cluster, the lowest redshift
  cluster in the sample at $z=0.064$, is dominated by two bright,
  probably interacting, galaxies. One of these galaxies has strong
  narrow H$\alpha$ emission and may contain a low luminosity AGN.  An
  8.7 mJy radio source is also present in the cluster center and coincides 
  with one of the bright central galaxies. The
  X-ray morphology, although from a low signal-to-noise image, may
  indicate some point-source contamination from within the cluster, as
  well as extended emission.  This cluster has been omitted from the
  statistically complete sample since it is in a ROSAT field targeted
  at another galaxy cluster.

\item[WARPJ1416.4$+$2315] This extended X-ray source originates in a $z=0.135$
  cluster of galaxies visible on both the sky survey plate and the deeper
  image.  A 4.1 mJy radio source (Table~\ref{tab:radio}) appears to be
  associated with the BCG.  Two of the three galaxies with redshifts (but not
  the BCG) show strong narrow H$\alpha$ emission, and one may be associated
  with weak X-ray emission. The X-ray flux from a QSO (HS1414+2330) behind the
  cluster also contributes to the quoted flux at about the 25\% level
  according to the WGACAT, \citep{wga}, but both of these sources are weak
  contaminants compared to the bright extended cluster emission. This a
  candidate for being a `fossil' group or cluster of galaxies. This cluster has
  been omitted from the statistically complete sample since it is in a field
  targeted at another galaxy cluster.

\item[WARPJ1418.5$+$2511] This massively extended, luminous
  (L$_{X,bol}\sim$10$^{45}$ erg s$^{-1}$) X-ray source originates in a $z=0.29$
  cluster of galaxies which can be seen both on the sky survey plate and the
  KPNO image.  Our cross-correlation with the NVSS reveals a 38.9 mJy radio
  source $65\arcsec$ from the X-ray centroid (Table~\ref{tab:radio}), which is
  associated with a likely cluster member.  We see no X-ray enhancement at the
  position of the radio source, so it is unlikely to contribute a major
  fraction of the X-ray emission.  The FIRST image reveals that this radio
  source may have a wide-angle-tail morphology, which is not at all atypical of
  cluster radio galaxies (e.g., \citealt{lo95}), although such sources are more
  commonly found in cluster cores. The Butcher-Oemler properties of this
  cluster have been studied by  \citet{fairley01}. The cluster has been
  previously discussed by \cite{Sto83} and other authors, as part of the
  Einstein Observatory Medium Sensitivity Survey (MSS, but note that this
  source is not in the later EMSS), and by \cite{nandra93} (it is only
  $\approx8\arcmin$ from the bright Seyfert 1 galaxy NGC 5548).  

\item[WARPJ1501.0$-$0824] This is one of two point-like X-ray sources
identified by our optical imaging program as a likely cluster of galaxies.  We
have not obtained optical spectroscopy since its X-ray flux is below the limit
for our statistically complete sample.  Aperture photometry of the BCG yields
$R=19.54 \pm 0.3$, which corresponds to a  redshift estimate of $z=0.51 \pm
0.07$ using the photometric estimation  method of \cite{vmf98}.  Our
cross-correlation with the NVSS (Table~\ref{tab:radio}) finds a resolved radio
source $23''$ from the X-ray source, which appears to be associated with a
likely cluster galaxy.  This indicates possible emission from an AGN; however,
the radio position is well outside the X-ray position's error circle, and in
addition the position of the radio source is well away from the optical center
of the cluster.  The field was later removed from consideration in the
statistically complete sample due to its high background. 

\item[WARPJ1515.5$+$4346] This cluster was included as part of a
  larger extremely diffuse source which was later resolved into three
  groupings.  The northern clump (now WARPJ1515.7$+$4352) appears
  point-like and spectroscopy reveals a broad-lined AGN at $z=0.524$.
  The southern clump is WARPJ1515.5$+$4346 at $z=0.135$ ($n_{z}$ = 3)
  but is just outside the 15$\arcmin$ radius from the ROSAT field
  center.  The identification of the central source is uncertain. It
  may be a cluster with a redshift $z=0.136 ~(n_z =1)$ or $z=0.237
  ~(n_z=2)$.  By itself, the central source is below our count rate
  cutoff.  Even if it is associated with WARPJ1515.5$+$4346, the
  centroid of the emission is still outside of the 15$\arcmin$ radius.
  Therefore, the cluster is not part of the statistically complete
  sample.  This source corresponds to VMF~169, which has a listed
  position about 4$\arcmin$ north, suggesting that the VMF source
  includes more of the diffuse emission.
 
  
\item[WARPJ1517.9$+$3127] This distant ($z=0.744, n_z = 2$), but
  luminous, cluster is just below our completeness threshold.  In
  later reprocessing with an updated VTP it failed to even make the
  count rate cutoff.  Therefore, it is not part of the statistically
  complete sample.
  
\item[WARPJ1537.7$+$1201] This X-ray source is associated with a group or poor
  cluster at $z=0.136$.  It has two X-ray emission components, which appear to
  be centered around the positions of the two brightest galaxies.  This source
  was decomposed into two components and was manually reinserted following
  inspection of the X-ray photon distribution  for the field.  The source is in
  a ROSAT field with high background and so the cluster has been excluded from
  our statistically complete sample.

\item[WARPJ1552.2$+$2013] One of the member galaxies, located at a minor peak
  in the X-ray surface brightness, has narrow, and possibly broad, H$\alpha$
  emission.  However, any contamination to the X-ray luminosity is at a small
  level.  This system is a candidate `fossil' group or poor cluster.

\item[WARPJ2038.4$-$0125] This extended X-ray source originates in one of the
  most distant ($z=0.679$) clusters of galaxies in the WARPS-I sample.  Two
  radio sources are located in the field which may be associated with cluster
  galaxies (Table~\ref{tab:radio}), but these are both well away from the X-ray
  emission contours.

\item[WARPJ2108.6$-$0507] This extended X-ray source is associated with a
  $z=0.222$ cluster which can be seen on both the sky survey plate and image. 
  A faint (3.7 mJy; Table~\ref{tab:radio}) radio source is located near the
  X-ray centroid, which does not appear to be associated with any of the
  brighter cluster galaxies as it is in a blank field according to our image. 
  The X-ray morphology is compact; there may be a cooling flow or point-source
  contamination, although none of the optical spectra for three objects in the
  cluster core suggest that they are the counterparts.

\item[WARPJ2108.7$-$0516] This extended X-ray source is associated with a
  $z=0.317$ cluster of galaxies.  A 9.6 mJy radio source
  (Table~\ref{tab:radio}) located $41\arcsec$ from the X-ray centroid, which
  does not correspond to the optical position of any galaxy visible on the CFHT
  image.  We have 10 spectroscopically confirmed galaxy members. The cluster
  has at least two components, both in X-ray morphology and galaxy velocities. 
  The x-ray source to the east is included in the quoted flux, but whether it
  is related to the cluster is not clear.

\item[WARPJ2146.0$+$0423] This massively extended X-ray source is associated
  with a $z=0.5324$ cluster of galaxies.  Our cross-correlation with the NVSS
  (Table~\ref{tab:radio}) reveals an unresolved radio source $30\arcsec$ from
  the X-ray centroid.  A likely cluster member is within the radio error
  circle, but the overlay of the optical and X-ray images does not indicate any
  enhancement of the X-ray emission near its position.  The source extent and
  lack of point source contamination is confirmed by an HRI observation.
  This cluster is also in the \cite{gho86} (GHO) sample. Oke (private
  communication) confirms the cluster redshift as $z=0.53$ (from 2
  redshifts) and also finds two (emission line) galaxies at z=0.40,
  suggesting a superposed foreground cluster may exist.  The
  Butcher-Oemler properties of this cluster have been studied by
  \cite{fairley01}, who also find evidence for a
  foreground group.  We have assumed all the X-ray flux originates in
  the $z=0.532$ system, since the X-ray emission is centered on the
  BCG of this system.  A small fraction of the quoted X-ray flux may
  arise in the bright nearby elliptical galaxy visible in the overlay.

\item[WARPJ2239.4$-$0547, WARPJ2239.6$-$0543] These two sources were
  found to be associated with the cluster Abell 2465.  The PSPC image
  shows two clearly extended X-ray sources, each centered on a
  well-defined cluster core.  We have obtained optical spectroscopy of
  objects in both clusters, which confirms them to be at slightly
  different redshifts ($z=0.2419$ versus $z=0.2427$ for six galaxies
  apiece).  We believe that these clusters may be in the process of
  merging.  The brightest galaxy of WARPJ2239.6-0543 is also an 8.4
  mJy radio source.  However, the morphology in an HRI image does not
  reveal any point source emission, and none of the optical spectra of
  galaxies in the cluster core show emission lines.
  
\item[WARPJ2239.5$-$0600] This X-ray source is associated with a
  compact group at $z=0.174$, and it falls just below the threshold
  for inclusion in our statistically complete sample at a flux of 6.30
  $\times 10^{-14}$ \fxunits.
  
\item[WARPJ2302.8$+$0843] Keck imaging and spectroscopy of this
  distant cluster ($z=0.722, n_z=1$) indicate that it may be contaminated by a
  BL Lac.  This is supported by the presence of a 25 mJy radio source
  very close to the cluster center.  The level of contamination is
  impossible to tell from the low signal-to-noise PSPC image.  A
  recent image of this source with {\it Chandra} shows that the X-ray
  emission is entirely extended, however (we will report on this image
  in a future paper).

\item[WARPJ2319.5$+$1226] We have only a good redshift for the BCG of this
  poor cluster or group ($z=0.125$) and have chosen to use the RIXOS value of
  z=0.124 ($n_z = 3$).

\item[WARPJ2320.7$+$1659] This clearly extended X-ray source is
  associated with a $z=0.499$ cluster of galaxies and a $z=1.81$ QSO
  at the x-ray peak.  Most of the flux may be from the QSO, but the
  level of contamination is impossible to tell from the low
  signal-to-noise PSPC image.  However, since the total flux is just
  above our flux limit, any small contribution from the QSO would
  reduce the cluster flux to a level below the flux limit, so we have
  excluded this source from our statistically complete sample. Our
  cross-correlation with the NVSS (Table~\ref{tab:radio}) also reveals
  a 2.8 mJy radio source $41\arcsec$ from the X-ray centroid, but no
  obvious X-ray enhancement is visible at this position.

\end{description}

\section{Discussion and Future Prospects}
\label{sec:prospects}

Here we have presented the identifications and cluster catalogs from
the first phase of the WARPS project, an extensive survey of deep
ROSAT pointings using a surface brightness sensitive algorithm.  In
doing so, we have also presented some comparisons of the results of
WARPS-I to other cluster surveys, which help place the the various
cluster detection algorithms and follow-up programs in perspective.  As
can be seen, the findings of WARPS are largely similar to those of
other cluster surveys as regards the space density of X-ray emitting
clusters.

The number of serendipitous X-ray surveys for clusters of galaxies now number
more than a dozen.  For example, the list of those based on ROSAT data includes
BCS \citep{bcsI,bcsII,bcsIII}, WARPS (Papers I, II), RDCS \citep{Ros98}, 
SHARC-Bright \citep{sharc2000}, SHARC-south \citep{Bur97,Col97}, CfA 160
deg$^2$ \citep{vmf98}, NEP \citep{nep}, NORAS \citep{noras}, REFLEX
\citep{reflexI}, MACS \citep{macs} and CIZA \citep{ciza}.  Each of these
surveys use different selection criteria and/or X-ray source-detection
algorithms, but largely similar optical follow-up programs.  Meanwhile, several
optical surveys have used moderate depth, large-area images of the sky combined
with algorithms that measure overdensities of galaxies (e.g., PDCS, 
\citealt{pdcsI, pdcsII}; RCS, \citealt{glye}). All of these surveys represent a
fertile ground for constraining cosmological parameters as well as learning
more about appropriate methods for finding massive clusters of galaxies and
their attendant selection effects.  However, precious few efforts (most notably
ROXS, \citealt{roxs}) are underway to understand the relationship of the
selection effects in optical and X-ray surveys.

Despite the different emphases inherent in each method, all of the new surveys
show no evidence for evolution at all but the most X-ray luminous clusters of
galaxies at redshifts up to $z=0.8$. The X-ray luminosity-temperature
relationship (Paper IV, \citealt{MS97}) at $z<0.8$, also shows no sign of
evolution.  Yet at $z \sim 1$, there are tantalizing hints of evolution in the
cluster population.  This evidence includes a preponderance of irregular X-ray
morphologies, optical cluster galaxy distributions and colors (Paper III,
\citealt{sta01}, although see Paper V for a notable counter-example), as well
as a suggestion that the space density of the most luminous 
($L_X > 5 \times 10^{44} {\rm ~erg ~s^{-1}}$) clusters decreases significantly 
at redshifts $z>0.5$ (\citealt{Ros98,vmfII}; although for a counterargument 
see Paper V).  Taken together, these lines of evidence might suggest that the 
epoch of
cluster formation is currently just out of reach of our current observational
tools, which are limited to finding clusters at $z<1.25$ by the limitations
inherent in observing in the optical (at $z=1.25$ the Ca break passes out of
the I band), even though simulations ({\it e.g.}, Paper I) show an ability to
detect X-ray luminous, extended sources at redshifts up to nearly $z=2$.

This presents future X-ray surveys ({\it e.g.}, {\it Chandra} or XMM-Newton)
with significant challenges, since one would expect them to contain large
numbers of $z>1$ clusters.  Indeed, as we pointed out in Papers III and V,
WARPS is sensitive enough to include clusters of $L_X \sim 10^{45} {\rm ~erg
~s^{-1}}$ out to redshifts of nearly 2. Even given the highly unclear (and very
poorly explored) state of galaxy evolution at $z\gtrsim 1$, it is obvious from
the current literature that, at $z\sim 0.8$, the space density and ICM
characteristics of clusters of galaxies show little difference from
those observed at lower redshifts.  It is therefore imperative that future
surveys for clusters of galaxies include near-infrared observations, which
should increase their window of sensitivity for identifying cluster
counterparts to X-ray sources out to $z\sim 4.5$ (i.e., when the H \& K break
passes out of the K band).  Of the current X-ray surveys, only RDCS has
included near-infrared observations in its follow-up program, and as a result,
RDCS is the only X-ray selected sample of clusters to contain $z>1$ objects
\citep{ros99, ros00}, albeit only a handful (4 according to \citealt{ros00} and
P. Rosati, private communication  from an incomplete survey of about 80\% of
RDCS blank-field X-ray sources).

The now overwhelming evidence of little or no significant evolution in
the space density of clusters at redshifts up to 0.8 is difficult to
reconcile with $\Omega_m=1$ cold dark matter cosmologies, which
predict significant evolution in the cluster LF at $z\gtrsim 0.3$ .
However, as \citet{bod01} show, a great deal of degeneracy remains in
the predictions of hierarchical models of structure formation at
$z>1$, and coincidentally, this is where the fewest such X-ray
selected clusters are known.  It is at these highest redshifts that
future X-ray and other surveys for clusters (notably
Sunyaev-Zel'dovich effect methods, see \citealt{xuwu01}) will have to
concentrate their efforts and where they will have the most impact.

\begin{acknowledgments}

E. S. P. acknowledges support from {\it Chandra} grant SAO1800627.  H. E.
acknowledges support from  NASA LTSA grant NAG5-8253.  C. A. S. acknowledges
support from NASA LTSA grant NAG5-3257  (PI: M. Donahue).  L. R. J.
acknowledges the support of the U. S. National Research Council and the UK
PPARC.
We acknowledge interesting conversations with and
emails from P. Rosati, C. Norman, M. Donahue, J. P. Henry, J. T. Stocke, R.
Mushotzky, M. Henriksen, A. Evrard and B. Mathiesen.

The WARPS team acknowledges the long-term support of many different time
allocation committees, including those at Kitt Peak National Observatory, Cerro
Tololo Inter-American Observatory and the University of Hawaii, for the
observations detailed in this paper.  Part of the observations used in this
paper were obtained at the MDM Observatory on Kitt Peak in Arizona.   Some of
the data presented herein were obtained at the W.M. Keck Observatory, which is
operated as a scientific partnership among  the California Institute of
Technology, the University of California and the National Aeronautics and Space
Administration.  The Keck Observatory was made  possible by the generous
financial support of the W.M. Keck Foundation.  This  research has made use of
data obtained through the High Energy Astrophysics Science Archive Research
Center Online Service, provided by the NASA/Goddard Space Flight Center. This
research has made use of the NASA/IPAC Extragalactic  Database (NED) which is
operated by the Jet Propulsion Laboratory, California  Institute of Technology,
under contract with the National Aeronautics and Space  Administration. This
research has also made use of the NRAO-VLA Sky Survey, which was carried out at
the NRAO Very Large Array.  NRAO is a facility of  Associated Universities,
Inc., and is a facility of the (USA) National Science Foundation.  Finally, we
wish to thank our referee, Dr. Luigi Guzzo, for helpful comments which improved
considerably this paper.

\end{acknowledgments}

\bibliography{apjmnemonic,warps}
\bibliographystyle{apj}

\clearpage

\clearpage

\figcaption{\label{fig:dssover} Overlays of the observed X-ray emission for the
  \extsources\ extended or marginally extended sources on digitized plates from
  the POSS-I, UKST, or POSS-II surveys.  Overlays for four additional X-ray
  sources which were found to be pointlike by VTP but were included as cluster
  candidates following our optical imaging program, are included at the end of
  the Figure.  Each image is $5\arcmin \times 5\arcmin$  with the standard
  orientation of north up and east to the left.  The x-rays were smoothed with
  a 45$\arcsec$ Gaussian.}
 
\figcaption{\label{fig:ccdover} Overlays of the observed X-ray
  emission for each cluster on an optical (R-band) CCD image.  Each
  image is 1.5 Mpc across at the cluster redshift (or 5$\arcmin$ if
  the redshift is unknown) with axes labelled in arc seconds from the
  VTP centroid with the standard orientation of north up and east to
  the left.  The x-rays were smoothed with a 45$\arcsec$ Gaussian.
  Radio sources are labelled R1, R2, etc. as in Table~\ref{tab:radio}
  and are marked with a circle the size of the positional error given
  in the Table (which may be too small to see).  Other sources (such
  as M stars) are also marked and labelled accordingly.}

\clearpage

\begin{deluxetable}{lrrrl}
\tablewidth{0pt}
\tabletypesize{\scriptsize}
\tablecaption{ROSAT PSPC pointings used in WARPS\label{tab:fields}}
\tablehead{\colhead{Sequence ID} & \colhead{Expo [sec.]} & \colhead{$\alpha$ (J2000)} & \colhead{$\delta$ (J2000)} & \colhead{Target Name}}
\startdata
rp150071n00 & 17912 & 14 17 59.9 & $+$5 08 23.9 &  NGC 5548          \\ 
rp200125n00 & 8614 & 16 55 33.6 & $-$08 23 24.1 &  LHS429            \\ 
rp200474n00 & 9372 & 23 18 45.5 & $+$2 36 00.1 &  KUV 2316+123      \\ 
rp200905n00 & 10327 & 15 13 33.5 & $+$8 34 11.8 &  SS BOOTIS         \\ 
rp200965n00 & 10328 & 15 14 47.9 & $+$4 01 47.8 &  PG1513+442        \\ 
rp201007n00 & 18199 & 21 09 19.1 & $-$13 14 24.1 &  LHS 65            \\ 
rp201018n00 & 19015 & 15 18 11.9 & $+$1 39 00.1 &  U CRB             \\ 
rp201019n00 & 10639 & 11 13 11.9 & $-$26 28 12.0 &  TT HYA            \\ 
rp201037n00 & 15533 & 13 56 09.5 & $+$5 55 12.0 &  ZZ BOO            \\ 
rp201045n00 & 28336 & 00 39 19.2 & $+$0 51 35.9 &  S AND             \\ 
rp201070n00 & 8301 & 13 29 47.9 & $+$1 06 00.0 &  HD117362          \\ 
rp201471n00 & 8373 & 13 11 52.7 & $+$7 52 48.1 &  HD 114710         \\ 
rp201505n00 & 28004 & 18 42 38.3 & $+$5 32 24.0 &  HD173524          \\ 
rp300003n00 & 24464 & 02 06 52.7 & $+$5 17 59.9 &  TT ARIETIS        \\ 
rp300021n00 & 24506 & 16 05 45.6 & $+$5 51 35.9 &  MS1603.6+2600     \\ 
rp300028n00 & 27116 & 13 48 55.2 & $+$7 57 35.8 &  PG1346+082        \\ 
rp300043n00 & 14804 & 01 41 00.0 & $-$67 53 24.1 &  BL HYI            \\ 
rp300079n00 & 49502 & 03 37 55.2 & $-$25 20 59.9 &  EXO 033319-2554.  \\ 
rp300180n00 & 45541 & 17 17 07.2 & $+$3 08 23.9 &  NGC 6341          \\ 
rp300218n00 & 20662 & 20 38 14.4 & $-$01 21 00.0 &  AE AQR            \\ 
rp300220n00 & 15603 & 23 16 02.3 & $-$05 27 00.0 &  RX J2316.1-0527   \\ 
rp300389n00 & 36716 & 21 07 55.1 & $-$05 16 12.1 &  RE2107-05         \\ 
rp400116n00 & 8303 & 12 59 59.9 & $+$2 40 11.9 &  PSR 1257+12       \\ 
rp400293n00 & 20080 & 22 54 19.1 & $+$9 03 36.1 &  GRB 910814        \\ 
rp400322n00 & 12424 & 00 19 50.4 & $+$1 56 59.9 &  RX J0019.8+2156   \\ 
rp600119n00 & 11993 & 15 06 31.2 & $+$5 46 12.0 &  NGC 5866          \\ 
rp600164n00 & 16752 & 12 58 59.9 & $+$4 51 36.0 &  NGC 4861          \\ 
rp600439n00 & 11233 & 23 20 31.1 & $+$7 13 48.0 &  III ZW 102        \\ 
rp600448n00 & 12808 & 14 34 52.8 & $+$8 40 48.0 &  MARK 474/NGC 568  \\ 
rp600462n00 & 14242 & 14 15 33.5 & $+$6 13 47.9 &  NGC 5529          \\ 
rp700006n00 & 12146 & 21 14 52.8 & $+$6 07 48.0 &  PG2112+059        \\ 
rp700026n00 & 11197 & 04 22 14.3 & $-$38 45 00.1 &  Q0420-388         \\ 
rp700027a01 & 17554 & 02 09 28.7 & $-$39 30 35.9 &  Q0207-398         \\ 
rp700061n00 & 25728 & 14 06 57.6 & $+$8 27 00.0 &  1404+286          \\ 
rp700096n00 & 8038 & 15 59 09.6 & $+$5 01 47.3 &  MKN493            \\ 
rp700101n00 & 24337 & 00 08 19.1 & $+$0 41 23.9 &  MKN335            \\ 
rp700108n00 & 13683 & 01 43 57.6 & $+$2 21 00.1 &  MKN573            \\ 
rp700114n00 & 9451 & 02 28 38.4 & $-$10 10 48.1 &  0226-1024         \\ 
rp700117n00 & 22456 & 14 06 43.2 & $+$4 11 23.6 &  3CR294            \\ 
rp700122n00 & 27863 & 14 15 45.6 & $+$1 29 42.0 &  Q1413+1143        \\ 
rp700331n00 & 23437 & 00 48 45.6 & $+$1 57 35.8 &  MKN 348           \\ 
rp700350n00 & 10028 & 02 35 07.1 & $-$04 01 48.0 &  0232-042          \\ 
rp700377n00 & 10718 & 00 46 14.4 & $+$1 04 11.8 &  Q0043+0048        \\ 
rp700423n00 & 18562 & 23 03 14.4 & $+$8 52 12.1 &  NGC 7469          \\ 
rp700432n00 & 13860 & 02 07 50.3 & $+$2 43 12.1 &  NAB 0205+024      \\ 
rp700527n00 & 8898 & 14 19 04.7 & $-$13 10 48.1 &  PG 1416-129       \\ 
rp700794n00 & 10830 & 01 11 28.8 & $-$38 04 48.0 &  NGC 424           \\ 
rp700796n00 & 9004 & 13 10 57.5 & $+$7 03 36.0 &  NGC 5005          \\ 
rp700797n00 & 9516 & 13 25 43.2 & $-$29 49 47.9 &  NGC 5135          \\ 
rp700908n00 & 9043 & 09 52 19.1 & $-$01 36 35.9 &  MKN 1239          \\ 
rp700920n00 & 12675 & 02 49 19.1 & $+$9 17 59.8 &  MKN372            \\ 
rp700976n00 & 11171 & 01 24 45.6 & $+$9 18 35.8 &  MS0122.1+0903     \\ 
rp700977n00 & 13951 & 02 08 38.3 & $+$5 23 24.1 &  MS0205.7+3509     \\ 
rp701000a01 & 27821 & 13 43 43.1 & $-$00 14 59.8 &  BJS864            \\ 
rp701018n00 & 11342 & 22 23 45.6 & $-$02 13 12.1 &  3C445             \\ 
rp701191n00 & 10395 & 22 12 59.9 & $-$17 10 12.1 &  RX J2213.0-1710   \\ 
rp701205n00 & 14428 & 23 43 31.2 & $-$14 55 12.1 &  MS2340.9-1511     \\ 
rp701206n00 & 9431 & 22 39 52.8 & $-$05 52 11.9 &  BR 2237-06        \\ 
rp701213n00 & 14664 & 15 52 09.5 & $+$0 05 59.9 &  3C326             \\ 
rp701250n00 & 18765 & 23 04 43.1 & $-$08 41 23.4 &  MCG-2-58-22       \\ 
rp701356n00 & 23554 & 02 37 12.0 & $-$52 15 36.0 &  ESO198-G24/ WWHO  \\ 
rp701373n00 & 15427 & 15 49 50.4 & $+$1 25 48.1 &  3C 324            \\ 
rp701403n00 & 11855 & 02 55 11.9 & $+$0 10 48.0 &  NGC 1144          \\ 
rp701405n00 & 13356 & 15 26 07.2 & $+$1 40 11.8 &  NGC 5929          \\ 
rp701411n00 & 22955 & 15 34 55.2 & $+$3 29 24.1 &  ARP220            \\ 
rp701420n00 & 9940 & 21 58 07.2 & $-$15 01 11.9 &  2155-15           \\ 
rp800003n00 & 28248 & 15 58 55.2 & $+$3 23 23.8 &  GC1556+335        \\ 
rp800150a01 & 12778 & 21 46 33.6 & $+$4 13 47.9 &  21 HOUR FIELD     \\ 
rp800401a01 & 11811 & 14 15 57.5 & $+$3 07 11.9 &  4C23.37           \\ 
rp800471n00 & 10639 & 04 30 45.6 & $+$0 24 36.1 &  NGC 1588          \\ 
rp800483n00 & 26671 & 00 22 52.7 & $+$4 24 35.5 &  CL0020            \\ 
rp900017n00 & 25815 & 04 40 55.1 & $-$16 30 36.1 &  EDS - PSPC        \\ 
rp900242n00 & 11620 & 04 14 16.8 & $-$12 44 24.0 &  NGC 1535          \\ 
rp900246n00 & 9152 & 04 11 40.7 & $-$17 28 48.1 &  ERID1             \\ 
rp900325n00 & 10322 & 13 17 43.1 & $+$4 54 00.0 &  G107+71           \\ 
rp900337n00 & 15618 & 22 25 48.0 & $-$04 57 00.0 &  2223-052          \\ 
rp900339n00 & 22143 & 22 53 57.5 & $+$6 08 59.9 &  2251+158          \\ 
rp900345n00 & 9346 & 21 27 55.2 & $+$2 58 12.0 &  G064-26           \\ 
rp900352n00 & 10383 & 02 17 19.1 & $-$17 45 35.9 &  G192-67           \\ 
rp900492n00 & 9609 & 03 48 23.9 & $-$12 10 11.9 &  ERID4             \\ 
\enddata
\end{deluxetable}


\begin{deluxetable}{llccrrrrrccccll}
\rotate
\tablewidth{0pt}
\tabletypesize{\scriptsize}
\tablecaption{WARPS-I Source Catalog\label{tab:highflux}}
\tablehead{\colhead{Name} & \colhead{ROSAT} & \colhead{ $\alpha$ } & \colhead{ $\delta$ } & \colhead{ cr$_{raw}$ } & \colhead{ cr$_{corr}$ } & \colhead{n$_{H}$ } & \colhead{Flux} & \colhead{Extent} & \colhead{Im} & \colhead{Sp} & \colhead{Id} & \colhead{Redshift/ } & \colhead{Other Name/ }\\
\colhead{ } & \colhead{Field} & \colhead{(J2000)} & \colhead{(J2000)} & \colhead{ [10$^{-3}$] } & \colhead{ [10$^{-3}$] } & \colhead{ [10$^{20}$] } & \colhead{ [10$^{-13}$] } & \colhead{ } & \colhead{ } & \colhead{ } & \colhead{ } & \colhead{Sp. Type} & \colhead{Notes}\\
\colhead{(1)} & \colhead{(2)} & \colhead{(3)} & \colhead{(4)} & \colhead{(5)} & \colhead{(6)} & \colhead{(7)} & \colhead{(8)} & \colhead{(9)} & \colhead{(10)} & \colhead{(11)} & \colhead{(12)} & \colhead{(13)} & \colhead{(14)}}
\startdata
WARPJ0008.4$+$2034 & rp700101n00 & 00 08 26.7 & $+$20 34 31 & 13.894 & 15.934 & 3.95 &    2.10 & 1.07 & \nodata & \nodata & qso & 0.389 & CRSS~J0008.4+2034 \\ 
WARPJ0008.8$+$2050 & rp700101n00 & 00 08 53.9 & $+$20 50 28 & 52.151 & 55.862 & 3.93 &    8.87 & 0.82 & \nodata & \nodata & star & M(e) & CRSS~J0008.9+2050 \\ 
WARPJ0019.8$+$2202 & rp400322n00 & 00 19 49.4 & $+$22 02 51 & 4.027 & 4.894 & 4.14 &    0.65 & 1.13 & R & Y & BLEND & \nodata & M~star+galaxy~IC~1541 \\ 
WARPJ0020.7$+$2205 & rp400322n00 & 00 20 43.3 & $+$22 05 36 & 6.143 & 11.651 & 4.14 &    1.54 & 1.81 & R & Y & \nodata & \nodata & \phn \\ 
WARPJ0038.2$+$3053 & rp201045n00 & 00 38 12.4 & $+$30 53 28 & 8.695 & 9.703 & 5.60 &    1.33 & 0.41 & \nodata & Y & gal & 0.018 & UGC~00381 \\ 
WARPJ0040.2$+$3054 & rp201045n00 & 00 40 12.1 & $+$30 54 31 & 10.434 & 11.536 & 5.60 &    1.58 & 0.98 & \nodata & Y & BLEND & \nodata & Sy1/QSO+NELG+Galaxy \\ 
WARPJ0045.2$+$0107 & rp700377n00 & 00 45 16.8 & $+$01 07 03 & 6.061 & 7.514 & 2.62 &    0.96 & 0.84 & \nodata & \nodata & \nodata & \nodata & \phn \\ 
WARPJ0111.0$-$3814 & rp700794n00 & 01 11 05.4 & $-$38 14 04 & 5.302 & 6.787 & 1.78 &    0.85 & 1.18 & BR & Y & \nodata & \nodata & \phn \\ 
WARPJ0111.6$-$3811 & rp700794n00 & 01 11 36.0 & $-$38 11 08 & 4.251 & 7.400 & 1.78 &    0.83 & 1.63 & BR & Y & CLG & 0.121 & VMF~009 \\ 
WARPJ0124.4$+$0921 & rp700976n00 & 01 24 25.2 & $+$09 21 40 & 5.358 & 6.166 & 4.70 &    0.98 & 1.05 & \nodata & \nodata & star & \nodata & GS~00614-00801 \\ 
WARPJ0124.7$+$0929 & rp700976n00 & 01 24 46.6 & $+$09 29 35 & 3.119 & 6.739 & 4.79 &    0.91 & 1.92 & R & \nodata & \nodata & \nodata & possible~blend,outskirts~of~NGC524 \\ 
WARPJ0124.8$+$0932 & rp700976n00 & 01 24 48.6 & $+$09 32 22 & 20.063 & 29.602 & 4.79 &    3.98 & 1.39 & \nodata & \nodata & gal & .0081 & NGC~524,SHARC \\ 
WARPJ0139.7$-$6749 & rp300043n00 & 01 39 43.8 & $-$67 49 12 & 18.353 & 20.043 & 2.35 &    2.55 & 0.33 & \nodata & \nodata & \nodata & \nodata & \phn \\ 
WARPJ0140.5$-$6805 & rp300043n00 & 01 40 31.3 & $-$68 05 09 & 12.986 & 15.570 & 2.70 &    1.99 & 0.26 & \nodata & \nodata & \nodata & \nodata & \phn \\ 
WARPJ0141.1$-$6756 & rp300043n00 & 01 41 10.2 & $-$67 56 38 & 4.633 & 5.431 & 2.35 &    0.69 & 1.06 & \nodata & \nodata & \nodata & \nodata & \phn \\ 
WARPJ0141.8$-$6759 & rp300043n00 & 01 41 51.0 & $-$67 59 08 & 7.181 & 8.542 & 2.61 &    1.09 & 0.81 & \nodata & \nodata & \nodata & \nodata & \phn \\ 
WARPJ0143.8$+$0224 & rp700108n00 & 01 43 48.7 & $+$02 24 55 & 4.838 & 5.720 & 2.91 &    0.74 & 1.08 & R & \nodata & \nodata & \nodata & \phn \\ 
WARPJ0144.5$+$0212 & rp700108n00 & 01 44 30.3 & $+$02 12 24 & 7.668 & 14.186 & 2.91 &    1.76 & 1.75 & R & Y & CLG & 0.165 & VMF~019 \\ 
WARPJ0144.5$+$0222 & rp700108n00 & 01 44 31.9 & $+$02 22 17 & 4.680 & 5.796 & 2.91 &    0.75 & 1.12 & \nodata & Y & AGN & 1.024 & \phn \\ 
WARPJ0206.1$+$1512 & rp300003n00 & 02 06 08.6 & $+$15 12 32 & 7.102 & 7.889 & 6.18 &    1.10 & 1.02 & R & Y & gal & 0.043 & IC~1777,RIXOS \\ 
WARPJ0206.3$+$1511 & rp300003n00 & 02 06 23.7 & $+$15 11 03 & 4.261 & 6.260 & 6.18 &    0.86 & 1.39 & R & Y & CLG & 0.251 & VMF~022,RIXOS \\ 
WARPJ0206.4$+$1529 & rp300003n00 & 02 06 24.6 & $+$15 29 04 & 3.750 & 4.956 & 6.18 &    0.69 & 0.74 & \nodata & Y & NELG & 0.301 & RIXOS \\ 
WARPJ0207.3$+$0231 & rp700432n00 & 02 07 22.5 & $+$02 31 31 & 13.613 & 15.194 & 3.11 &    1.97 & 0.83 & \nodata & \nodata & agn & 0.673 & MS~0204.8+0217 \\ 
WARPJ0207.3$+$1509 & rp300003n00 & 02 07 21.9 & $+$15 09 38 & 4.056 & 4.752 & 6.18 &    0.66 & 0.55 & R & Y & AGN & 0.149 & RIXOS~F246~040,RIXOS \\ 
WARPJ0207.9$+$3518 & rp700977n00 & 02 07 57.9 & $+$35 18 34 & 5.457 & 6.063 & 6.26 &    0.84 & 1.37 & \nodata & Y & AGN & 0.188 & \phn \\ 
WARPJ0210.4$-$3929 & rp700027a01 & 02 10 26.7 & $-$39 29 27 & 4.004 & 7.434 & 1.43 &    0.90 & 1.76 & R & Y & CLG & 0.273: & VMF~025 \\ 
WARPJ0216.5$-$1747 & rp900352n00 & 02 16 32.8 & $-$17 47 11 & 7.962 & 13.422 & 2.98 &    1.75 & 1.60 & R & Y & CLG & 0.578 & \phn \\ 
WARPJ0217.4$-$1800 & rp900352n00 & 02 17 26.5 & $-$18 00 10 & 13.010 & 14.973 & 2.98 &    1.93 & 1.09 & R & \nodata & blend & \nodata & SHARC,agn+? \\ 
WARPJ0217.7$-$1750 & rp900352n00 & 02 17 46.5 & $-$17 50 25 & 10.572 & 11.342 & 2.56 &    1.45 & 1.03 & \nodata & \nodata & \nodata & \nodata & \phn \\ 
WARPJ0227.7$-$1014 & rp700114n00 & 02 27 42.8 & $-$10 14 17 & 4.894 & 7.793 & 2.51 &    0.99 & 1.49 & R & \nodata & \nodata & \nodata & \phn \\ 
WARPJ0228.1$-$1005 & rp700114n00 & 02 28 11.7 & $-$10 05 30 & 14.114 & 25.960 & 2.51 &    3.23 & 1.75 & R & Y & CLG & 0.149 & VMF~026 \\ 
WARPJ0229.2$-$1009 & rp700114n00 & 02 29 13.7 & $-$10 09 37 & 18.266 & 19.687 & 2.69 &    2.52 & 0.45 & \nodata & \nodata & \nodata & \nodata & \phn \\ 
WARPJ0229.5$-$1005 & rp700114n00 & 02 29 35.4 & $-$10 05 55 & 11.006 & 12.622 & 2.69 &    1.62 & 1.06 & \nodata & Y & \nodata & \nodata & \phn \\ 
WARPJ0235.3$-$0347 & rp700350n00 & 02 35 19.4 & $-$03 47 11 & 27.030 & 28.997 & 2.70 &    3.71 & 0.67 & \nodata & \nodata & agn & 0.376 & MS~0232.8-0400 \\ 
WARPJ0235.5$-$0351 & rp700350n00 & 02 35 34.9 & $-$03 51 48 & 6.578 & 7.243 & 2.70 &    0.93 & 1.04 & \nodata & \nodata & \nodata & \nodata & \phn \\ 
WARPJ0236.1$-$5219 & rp701356n00 & 02 36 11.8 & $-$52 19 13 & 5.783 & 6.383 & 3.08 &    0.83 & 1.09 & R & \nodata & \nodata & \nodata & \phn \\ 
WARPJ0236.5$-$5227 & rp701356n00 & 02 36 30.1 & $-$52 27 01 & 5.372 & 6.482 & 3.08 &    0.84 & 0.64 & \nodata & \nodata & \nodata & \nodata & \phn \\ 
WARPJ0236.8$-$5203 & rp701356n00 & 02 36 51.6 & $-$52 03 00 & 137.130 & 145.207 & 3.07 &   23.04 & 1.06 & \nodata & \nodata & star & M2 & EXO~0235.2-5216 \\ 
WARPJ0237.0$-$5223 & rp701356n00 & 02 37 01.9 & $-$52 23 47 & 56.271 & 59.694 & 3.07 &    7.71 & 0.12 & \nodata & \nodata & agn & 0.113 & EXO~0235.3-5236 \\ 
WARPJ0237.2$-$5227 & rp701356n00 & 02 37 13.6 & $-$52 27 32 & 16.112 & 17.393 & 3.07 &    2.25 & 0.51 & \nodata & \nodata & \nodata & \nodata & \phn \\ 
WARPJ0238.0$-$5224 & rp701356n00 & 02 38 01.2 & $-$52 24 41 & 30.397 & 40.674 & 3.07 &    5.18 & 1.28 & R & Y & CLG & 0.135 & A3038,VMF~028,SHARC \\ 
WARPJ0250.0$+$1908 & rp700920n00 & 02 50 03.5 & $+$19 08 00 & 6.852 & 11.547 & 9.87 &    1.68 & 1.56 & BR & Y & CLG & 0.122 & SHARC \\ 
WARPJ0255.0$+$0017 & rp701403n00 & 02 55 01.3 & $+$00 17 49 & 21.072 & 22.680 & 6.48 &    3.17 & 1.04 & R & \nodata & \nodata & \nodata & \phn \\ 
WARPJ0255.0$+$0025 & rp701403n00 & 02 55 05.7 & $+$00 25 26 & 33.471 & 35.758 & 6.64 &    5.02 & 0.99 & \nodata & Y & qso & 0.354 & US~3333 \\ 
WARPJ0256.0$+$0015 & rp701403n00 & 02 56 04.0 & $+$00 15 55 & 5.431 & 7.821 & 6.64 &    1.10 & 1.35 & R & Y & BLEND & \nodata & NELG~(z=0.14)~+~? \\ 
WARPJ0337.4$-$2518 & rp300079n00 & 03 37 28.6 & $-$25 18 48 & 4.799 & 6.046 & 1.55 &    0.75 & 1.16 & R & \nodata & \nodata & \nodata & \phn \\ 
WARPJ0337.5$-$2518 & rp300079n00 & 03 37 35.2 & $-$25 18 10 & 12.108 & 14.021 & 1.55 &    1.75 & 1.08 & R & \nodata & \nodata & \nodata & SHARC \\ 
WARPJ0337.6$-$2523 & rp300079n00 & 03 37 37.9 & $-$25 23 23 & 5.759 & 6.545 & 1.55 &    0.82 & 1.05 & R & \nodata & \nodata & \nodata & \phn \\ 
WARPJ0338.4$-$2524 & rp300079n00 & 03 38 27.6 & $-$25 23 59 & 5.152 & 5.746 & 1.55 &    0.72 & 0.86 & \nodata & \nodata & \nodata & \nodata & \phn \\ 
WARPJ0338.6$-$2529 & rp300079n00 & 03 38 36.0 & $-$25 29 10 & 8.171 & 9.553 & 1.55 &    1.19 & 0.87 & \nodata & \nodata & \nodata & \nodata & \phn \\ 
WARPJ0347.6$-$1217 & rp900492n00 & 03 47 40.7 & $-$12 16 59 & 3.651 & 5.040 & 3.84 &    0.66 & 1.22 & R & \nodata & \nodata & \nodata & \phn \\ 
WARPJ0347.8$-$1217 & rp900492n00 & 03 47 50.4 & $-$12 17 01 & 3.948 & 5.057 & 3.84 &    0.67 & 1.08 & R & Y & \nodata & \nodata & \phn \\ 
WARPJ0348.1$-$1201 & rp900492n00 & 03 48 08.7 & $-$12 01 34 & 5.051 & 7.398 & 3.80 &    0.97 & 1.38 & R & Y & CLG & 0.488 & \phn \\ 
WARPJ0349.3$-$1205 & rp900492n00 & 03 49 18.7 & $-$12 05 43 & 6.414 & 7.852 & 3.80 &    1.03 & 0.94 & \nodata & \nodata & \nodata & \nodata & \phn \\ 
WARPJ0411.5$-$1717 & rp900246n00 & 04 11 34.7 & $-$17 17 25 & 5.909 & 6.875 & 2.52 &    0.88 & 1.05 & \nodata & Y & AGN & \nodata & \phn \\ 
WARPJ0411.9$-$1728 & rp900246n00 & 04 11 58.4 & $-$17 28 04 & 8.347 & 9.800 & 2.49 &    1.25 & 1.10 & \nodata & \nodata & \nodata & \nodata & \phn \\ 
WARPJ0421.7$-$3847 & rp700026n00 & 04 21 47.1 & $-$38 47 47 & 4.409 & 6.044 & 1.90 &    0.76 & 1.26 & R & \nodata & \nodata & \nodata & \phn \\ 
WARPJ0421.8$-$3833 & rp700026n00 & 04 21 52.5 & $-$38 33 00 & 3.777 & 5.587 & 1.90 &    0.70 & 0.53 & \nodata & \nodata & \nodata & \nodata & \phn \\ 
WARPJ0422.3$-$3832 & rp700026n00 & 04 22 20.8 & $-$38 32 20 & 4.843 & 6.719 & 1.90 &    0.84 & 6.79 & \nodata & \nodata & qso & 0.482 & MS~0420.6-3839 \\ 
WARPJ0440.8$-$1635 & rp900017n00 & 04 40 53.2 & $-$16 35 22 & 9.930 & 10.747 & 4.69 &    1.71 & 0.15 & R & \nodata & star & G0 & 1E~0438.6-1641 \\ 
WARPJ0441.0$-$1616 & rp900017n00 & 04 41 05.2 & $-$16 16 08 & 9.427 & 10.916 & 4.69 &    1.47 & 1.03 & R & \nodata & \nodata & \nodata & \phn \\ 
WARPJ0951.7$-$0128 & rp700908n00 & 09 51 47.0 & $-$01 28 30 & 2.995 & 5.536 & 3.84 &    0.73 & 1.73 & R & Y & CLG & 0.568 & VMF~076 \\ 
WARPJ0952.1$-$0148 & rp700908n00 & 09 52 08.5 & $-$01 48 20 & 3.777 & 6.230 & 3.84 &    0.82 & 1.56 & R & \nodata & \nodata & \nodata & VMF~077 \\ 
WARPJ0952.5$-$0137 & rp700908n00 & 09 52 34.8 & $-$01 37 53 & 4.047 & 5.307 & 3.84 &    0.70 & 1.22 & \nodata & \nodata & \nodata & \nodata & \phn \\ 
WARPJ0952.7$-$0130 & rp700908n00 & 09 52 43.3 & $-$01 30 47 & 9.761 & 11.503 & 3.84 &    1.51 & 1.10 & \nodata & Y & BLEND & \nodata & QSO~(z=1.26)+K~star \\ 
WARPJ1112.8$-$2619 & rp201019n00 & 11 12 50.5 & $-$26 19 07 & 8.632 & 10.461 & 5.47 &    1.43 & 1.12 & IR & \nodata & \nodata & \nodata & \phn \\ 
WARPJ1113.0$-$2615 & rp201019n00 & 11 13 05.7 & $-$26 15 36 & 4.096 & 5.424 & 5.47 &    0.75 & 1.21 & IR & Y & CLG & 0.725 & \phn \\ 
WARPJ1259.1$+$3501 & rp600164n00 & 12 59 09.2 & $+$35 01 15 & 5.418 & 6.292 & 1.22 &    0.78 & 0.83 & \nodata & \nodata & \nodata & \nodata & \phn \\ 
WARPJ1259.5$+$1226 & rp400116n00 & 12 59 33.3 & $+$12 26 47 & 4.049 & 8.065 & 2.27 &    1.02 & 1.82 & R & \nodata & \nodata & \nodata & \phn \\ 
WARPJ1300.0$+$3450 & rp600164n00 & 13 00 03.0 & $+$34 50 40 & 3.658 & 5.330 & 1.20 &    0.66 & 1.04 & \nodata & \nodata & \nodata & \nodata & FIRST \\ 
WARPJ1316.4$+$4454 & rp900325n00 & 13 16 25.7 & $+$44 54 18 & 21.064 & 23.867 & 1.45 &    2.97 & 0.93 & \nodata & \nodata & \nodata & \nodata & \phn \\ 
WARPJ1317.4$+$4456 & rp900325n00 & 13 17 28.3 & $+$44 56 40 & 5.406 & 6.970 & 1.54 &    0.87 & 1.19 & \nodata & \nodata & \nodata & \nodata & \phn \\ 
WARPJ1318.6$+$4500 & rp900325n00 & 13 18 39.9 & $+$45 00 48 & 4.705 & 6.040 & 1.78 &    0.76 & 1.19 & R & Y & BLEND & 0.173 & AGN~(z=0.173:)~+~? \\ 
WARPJ1328.9$+$0112 & rp201070n00 & 13 28 56.1 & $+$01 12 15 & 4.356 & 6.759 & 1.85 &    0.85 & 3.15 & R & \nodata & \nodata & \nodata & \phn \\ 
WARPJ1343.0$-$0013 & rp701000a01 & 13 43 02.6 & $-$00 13 01 & 7.140 & 8.625 & 1.91 &    1.08 & 3.00 & \nodata & \nodata & \nodata & \nodata & \phn \\ 
WARPJ1348.5$+$0757 & rp300028n00 & 13 48 34.9 & $+$07 57 37 & 7.305 & 8.087 & 2.00 &    1.28 & 0.08 & \nodata & \nodata & star & F5 & HD~120317 \\ 
WARPJ1356.1$+$2606 & rp201037n00 & 13 56 09.6 & $+$26 06 41 & 4.794 & 6.227 & 1.30 &    0.77 & 1.20 & R & Y & BLEND & 1.320 & QSO~(z=1.32)~+~? \\ 
WARPJ1356.3$+$2552 & rp201037n00 & 13 56 22.0 & $+$25 52 49 & 4.670 & 5.442 & 1.34 &    0.67 & 1.06 & R & \nodata & \nodata & \nodata & \phn \\ 
WARPJ1406.2$+$2830 & rp700061n00 & 14 06 16.1 & $+$28 30 55 & 4.739 & 7.065 & 1.46 &    0.87 & 1.39 & R & Y & CLG & 0.546 & VMF~153 \\ 
WARPJ1406.9$+$2834 & rp700061n00 & 14 06 55.4 & $+$28 34 25 & 11.321 & 19.056 & 1.46 &    2.25 & 1.58 & BR & Y & CLG & 0.117 & VMF~154,SHARC \\ 
WARPJ1407.3$+$2818 & rp700061n00 & 14 07 19.5 & $+$28 18 19 & 14.814 & 16.501 & 1.46 &    2.05 & 0.48 & \nodata & \nodata & qso & 1.121 & CRSS~J1407.3+2818 \\ 
WARPJ1407.7$+$2830 & rp700061n00 & 14 07 45.5 & $+$28 30 31 & 26.856 & 30.942 & 1.46 &    3.85 & 0.81 & R & \nodata & qso & 0.642 & CRSS~J1407.7+2830,FIRST \\ 
WARPJ1415.1$+$3612 & rp600462n00 & 14 15 11.1 & $+$36 12 03 & 6.295 & 9.132 & 1.14 &    1.13 & 1.36 & R & \nodata & CLG & [0.7] & FIRST \\ 
WARPJ1415.2$+$1119 & rp700122n00 & 14 15 16.8 & $+$11 19 30 & 2.855 & 4.671 & 1.80 &    0.74 & 1.40 & R & \nodata & star & G & CRSS~J1415.2+1119 \\ 
WARPJ1415.2$+$3608 & rp600462n00 & 14 15 12.4 & $+$36 08 09 & 14.248 & 15.377 & 1.14 &    1.90 & 0.99 & \nodata & \nodata & \nodata & \nodata & \phn \\ 
WARPJ1415.5$+$1131 & rp700122n00 & 14 15 31.8 & $+$11 31 57 & 10.341 & 11.810 & 1.80 &    1.48 & 1.06 & \nodata & \nodata & qso & 0.256 & CRSS~J1415.5+1131 \\ 
WARPJ1415.6$+$1124 & rp700122n00 & 14 15 40.3 & $+$11 24 10 & 3.865 & 4.388 & 1.80 &    0.70 & 0.77 & \nodata & \nodata & star & M(e) & CRSS~J1415.6+1124 \\ 
WARPJ1417.3$+$2505 & rp150071n00 & 14 17 22.1 & $+$25 05 22 & 13.475 & 14.796 & 1.58 &    1.84 & 0.44 & \nodata & \nodata & \nodata & \nodata & \phn \\ 
WARPJ1417.3$+$2513 & rp150071n00 & 14 17 23.2 & $+$25 13 41 & 5.259 & 6.691 & 1.58 &    0.83 & 0.21 & \nodata & \nodata & qso & 0.56 & 2E~1416.6+2523 \\ 
WARPJ1418.5$+$2511 & rp150071n00 & 14 18 31.8 & $+$25 10 59 & 39.210 & 48.915 & 1.93 &    6.20 & 1.18 & R & Y & CLG & 0.296 & VMF~159,SHARC \\ 
WARPJ1418.6$-$1306 & rp700527n00 & 14 18 38.1 & $-$13 06 36 & 5.468 & 6.230 & 6.29 &    0.87 & 1.02 & \nodata & \nodata & \nodata & \nodata & \phn \\ 
WARPJ1418.9$+$2509 & rp150071n00 & 14 18 58.2 & $+$25 09 49 & 7.580 & 8.535 & 1.93 &    1.07 & 0.73 & \nodata & \nodata & qso & 0.674 & [HB89]~1416+254 \\ 
WARPJ1419.5$-$1321 & rp700527n00 & 14 19 33.2 & $-$13 21 13 & 4.541 & 6.819 & 7.11 &    0.97 & 1.38 & R & \nodata & \nodata & \nodata & \phn \\ 
WARPJ1419.6$-$1316 & rp700527n00 & 14 19 39.3 & $-$13 16 35 & 6.108 & 7.011 & 7.11 &    1.00 & 1.05 & R & \nodata & \nodata & \nodata & \phn \\ 
WARPJ1435.4$+$4845 & rp600448n00 & 14 35 29.0 & $+$48 45 17 & 4.407 & 6.475 & 2.05 &    0.82 & 1.38 & \nodata & \nodata & gal & 0.0072 & NGC~5689 \\ 
WARPJ1505.4$+$5540 & rp600119n00 & 15 05 24.1 & $+$55 40 34 & 4.584 & 5.282 & 1.47 &    0.84 & 0.81 & \nodata & \nodata & star & F8 & HD~134023 \\ 
WARPJ1505.7$+$5549 & rp600119n00 & 15 05 43.9 & $+$55 49 33 & 8.573 & 10.047 & 1.47 &    1.25 & 1.08 & \nodata & \nodata & \nodata & \nodata & \phn \\ 
WARPJ1506.8$+$5532 & rp600119n00 & 15 06 51.1 & $+$55 32 06 & 5.806 & 7.474 & 1.47 &    0.93 & 0.84 & \nodata & \nodata & \nodata & \nodata & \phn \\ 
WARPJ1513.5$+$3846 & rp200905n00 & 15 13 32.3 & $+$38 46 00 & 6.307 & 7.328 & 1.39 &    0.91 & 0.41 & R & \nodata & \nodata & \nodata & \phn \\ 
WARPJ1513.8$+$4400 & rp200965n00 & 15 13 48.9 & $+$44 00 06 & 5.331 & 6.599 & 1.95 &    0.83 & 1.16 & R & \nodata & \nodata & \nodata & \phn \\ 
WARPJ1515.6$+$4411 & rp200965n00 & 15 15 39.1 & $+$44 11 29 & 9.071 & 10.275 & 2.04 &    1.29 & 1.04 & R & \nodata & \nodata & \nodata & \phn \\ 
WARPJ1517.0$+$3140 & rp201018n00 & 15 17 03.6 & $+$31 40 45 & 5.334 & 6.127 & 1.91 &    0.77 & 0.97 & R & \nodata & \nodata & \nodata & SHARC \\ 
WARPJ1517.2$+$3141 & rp201018n00 & 15 17 13.0 & $+$31 41 25 & 5.260 & 6.160 & 1.91 &    0.98 & 0.67 & R & \nodata & star & G5 & PPM~78477,FIRST \\ 
WARPJ1519.3$+$3139 & rp201018n00 & 15 19 19.6 & $+$31 39 07 & 5.401 & 6.068 & 1.90 &    0.76 & 0.04 & R & \nodata & \nodata & \nodata & \phn \\ 
WARPJ1524.9$+$4143 & rp701405n00 & 15 24 59.9 & $+$41 43 37 & 9.035 & 10.256 & 2.08 &    1.29 & 1.05 & \nodata & \nodata & \nodata & \nodata & FIRST \\ 
WARPJ1525.1$+$4145 & rp701405n00 & 15 25 10.2 & $+$41 45 58 & 4.360 & 5.195 & 2.08 &    0.66 & 1.10 & \nodata & \nodata & \nodata & \nodata & \phn \\ 
WARPJ1525.8$+$4139 & rp701405n00 & 15 25 49.1 & $+$41 39 28 & 4.134 & 5.820 & 2.08 &    0.73 & 1.31 & \nodata & \nodata & BLEND & \nodata & \phn \\ 
WARPJ1525.9$+$4148 & rp701405n00 & 15 25 58.2 & $+$41 48 36 & 7.995 & 9.170 & 2.08 &    1.16 & 1.07 & \nodata & \nodata & \nodata & \nodata & \phn \\ 
WARPJ1527.1$+$4133 & rp701405n00 & 15 27 08.8 & $+$41 33 12 & 5.721 & 7.750 & 2.08 &    1.23 & 0.31 & \nodata & \nodata & star & G5 & HD~137896,FIRST \\ 
WARPJ1548.8$+$2120 & rp701373n00 & 15 48 51.9 & $+$21 20 27 & 7.421 & 8.665 & 4.30 &    1.15 & 1.05 & \nodata & \nodata & \nodata & \nodata & \phn \\ 
WARPJ1549.7$+$2114 & rp701373n00 & 15 49 43.2 & $+$21 14 02 & 16.453 & 18.612 & 4.30 &    2.48 & 1.12 & \nodata & \nodata & \nodata & \nodata & \phn \\ 
WARPJ1550.2$+$2124 & rp701373n00 & 15 50 15.1 & $+$21 24 30 & 4.658 & 5.046 & 4.30 &    0.67 & 1.03 & \nodata & \nodata & \nodata & \nodata & \phn \\ 
WARPJ1552.0$+$2013 & rp701213n00 & 15 52 02.5 & $+$20 13 58 & 70.435 & 80.424 & 3.70 &   10.55 & 1.07 & R & \nodata & qso & 0.250 & [HB89]~1549+203 \\ 
WARPJ1552.2$+$2013 & rp701213n00 & 15 52 12.5 & $+$20 13 32 & 11.883 & 25.704 & 3.70 &    3.29 & 2.04 & BR & Y & CLG & 0.135 & VMF~175,SHARC \\ 
WARPJ1552.4$+$2007 & rp701213n00 & 15 52 29.8 & $+$20 07 18 & 24.633 & 28.092 & 3.70 &    3.69 & 1.06 & R & \nodata & \nodata & \nodata & RadioS \\ 
WARPJ1558.1$+$3500 & rp700096n00 & 15 58 09.7 & $+$35 00 52 & 5.429 & 6.994 & 2.03 &    0.88 & 1.94 & R & \nodata & \nodata & \nodata & \phn \\ 
WARPJ1558.7$+$3511 & rp700096n00 & 15 58 43.6 & $+$35 11 19 & 4.126 & 6.575 & 1.86 &    0.83 & 1.45 & R & \nodata & \nodata & \nodata & \phn \\ 
WARPJ1655.4$-$0820 & rp200125n00 & 16 55 29.3 & $-$08 20 00 & 1268.800 & 1344.763 & 13.38 &  213.40 & 1.28 & \nodata & \nodata & star & M3Ve & V1054~Oph \\ 
WARPJ1656.1$-$0819 & rp200125n00 & 16 56 11.7 & $-$08 19 10 & 4.996 & 5.983 & 13.38 &    0.97 & 0.01 & \nodata & \nodata & \nodata & \nodata & \phn \\ 
WARPJ1716.5$+$4302 & rp300180n00 & 17 16 32.6 & $+$43 02 27 & 53.605 & 57.084 & 2.19 &    9.06 & 0.67 & \nodata & Y & STAR & G/K & in~NGC~6341 \\ 
WARPJ1841.5$+$5531 & rp201505n00 & 18 41 31.1 & $+$55 31 17 & 4.598 & 5.269 & 4.95 &    0.71 & 1.68 & \nodata & Y & AGN & \nodata & \phn \\ 
WARPJ1841.8$+$5540 & rp201505n00 & 18 41 50.9 & $+$55 40 48 & 6.510 & 7.783 & 4.95 &    1.05 & 1.08 & \nodata & Y & \nodata & \nodata & RadioS \\ 
WARPJ1842.7$+$5529 & rp201505n00 & 18 42 43.9 & $+$55 29 22 & 31.186 & 33.249 & 4.95 &    4.49 & 1.09 & \nodata & Y & \nodata & \nodata & RadioS \\ 
WARPJ1843.4$+$5527 & rp201505n00 & 18 43 26.7 & $+$55 27 44 & 4.357 & 5.708 & 4.95 &    0.77 & 1.23 & \nodata & Y & BLEND & \nodata & \phn \\ 
WARPJ2037.6$-$0125 & rp300218n00 & 20 37 36.7 & $-$01 25 28 & 17.696 & 19.070 & 7.02 &    2.70 & 1.02 & \nodata & Y & AGN & 0.132 & \phn \\ 
WARPJ2038.3$-$0106 & rp300218n00 & 20 38 20.2 & $-$01 06 19 & 9.960 & 11.400 & 7.02 &    1.81 & 4.85 & \nodata & \nodata & star & G8/K0 & HD~196574 \\ 
WARPJ2038.4$-$0125 & rp300218n00 & 20 38 29.6 & $-$01 25 11 & 2.654 & 4.842 & 7.02 &    0.70 & 1.71 & BR & Y & CLG & 0.679 & \phn \\ 
WARPJ2038.7$-$0114 & rp300218n00 & 20 38 47.9 & $-$01 14 24 & 15.851 & 17.455 & 7.02 &    2.47 & 1.06 & \nodata & \nodata & \nodata & \nodata & \phn \\ 
WARPJ2107.6$-$0512 & rp300389n00 & 21 07 36.6 & $-$05 12 11 & 11.022 & 12.464 & 5.57 &    1.71 & 1.05 & \nodata & \nodata & \nodata & \nodata & \phn \\ 
WARPJ2108.6$-$1316 & rp201007n00 & 21 08 40.7 & $-$13 16 46 & 7.547 & 8.599 & 4.09 &    1.14 & 1.03 & \nodata & \nodata & \nodata & \nodata & \phn \\ 
WARPJ2108.8$-$0516 & rp300389n00 & 21 08 51.0 & $-$05 16 32 & 9.330 & 13.706 & 5.57 &    1.89 & 1.40 & R & Y & CLG & 0.317 & VMF~200 \\ 
WARPJ2109.3$-$1310 & rp201007n00 & 21 09 21.1 & $-$13 10 58 & 3.308 & 4.096 & 4.09 &    0.65 & 1.13 & \nodata & \nodata & star & F8 & HD~358246 \\ 
WARPJ2115.2$+$0608 & rp700006n00 & 21 15 16.4 & $+$06 08 43 & 10.952 & 12.826 & 6.62 &    1.80 & 1.09 & \nodata & Y & qso & 0.398 & [HB89]~2112+059 \\ 
WARPJ2126.9$+$1257 & rp900345n00 & 21 26 54.5 & $+$12 57 39 & 8.179 & 10.032 & 6.38 &    1.40 & 1.10 & \nodata & Y & AGN & 0.82 & \phn \\ 
WARPJ2127.0$+$1251 & rp900345n00 & 21 27 02.4 & $+$12 51 10 & 5.202 & 6.743 & 6.38 &    0.94 & 0.61 & \nodata & Y & AGN & 0.22 & \phn \\ 
WARPJ2158.4$-$1450 & rp701420n00 & 21 58 28.7 & $-$14 50 34 & 4.140 & 7.238 & 3.75 &    0.95 & 1.63 & R & Y & BLEND & 1.075 & QSO~(z=1.08)~+~? \\ 
WARPJ2158.9$-$1452 & rp701420n00 & 21 58 55.2 & $-$14 52 53 & 3.951 & 5.128 & 3.75 &    0.67 & 1.11 & R & \nodata & \nodata & \nodata & \phn \\ 
WARPJ2223.5$-$0218 & rp701018n00 & 22 23 32.8 & $-$02 18 04 & 9.678 & 11.105 & 4.95 &    1.76 & 1.07 & \nodata & \nodata & star & F8 & HD~212337 \\ 
WARPJ2223.8$-$0206 & rp701018n00 & 22 23 48.6 & $-$02 06 19 & 19.832 & 26.063 & 4.95 &    3.52 & 1.24 & R & \nodata & agn & 0.056 & 3C~445,SHARC,FIRST \\ 
WARPJ2224.3$-$0218 & rp701018n00 & 22 24 22.7 & $-$02 18 33 & 5.917 & 7.770 & 4.95 &    1.05 & 1.23 & R & \nodata & \nodata & \nodata & \phn \\ 
WARPJ2225.0$-$0457 & rp900337n00 & 22 25 00.5 & $-$04 57 51 & 5.670 & 6.554 & 5.07 &    0.89 & 0.77 & R & \nodata & \nodata & \nodata & \phn \\ 
WARPJ2239.4$-$0547 & rp701206n00 & 22 39 24.8 & $-$05 47 10 & 15.813 & 21.800 & 4.01 &    2.88 & 1.30 & BR & Y & CLG & 0.242 & A2465S,VMF~208 \\ 
WARPJ2239.6$-$0543 & rp701206n00 & 22 39 39.6 & $-$05 43 15 & 17.600 & 30.484 & 4.01 &    4.04 & 1.65 & BR & Y & CLG & 0.243 & A2465N,VMF~210 \\ 
WARPJ2253.5$+$2852 & rp400293n00 & 22 53 35.2 & $+$28 52 11 & 5.336 & 6.187 & 5.60 &    0.85 & 0.77 & R & \nodata & \nodata & \nodata & \phn \\ 
WARPJ2253.6$+$2913 & rp400293n00 & 22 53 37.6 & $+$29 13 13 & 21.089 & 22.997 & 5.68 &    3.16 & 0.74 & \nodata & \nodata & \nodata & \nodata & \phn \\ 
WARPJ2254.1$+$1606 & rp900339n00 & 22 54 10.9 & $+$16 06 56 & 9.188 & 9.921 & 6.54 &    1.39 & 1.04 & \nodata & \nodata & \nodata & \nodata & \phn \\ 
WARPJ2254.8$+$1604 & rp900339n00 & 22 54 49.8 & $+$16 04 21 & 10.281 & 11.276 & 6.54 &    1.58 & 0.99 & \nodata & Y & \nodata & \nodata & \phn \\ 
WARPJ2255.4$+$2905 & rp400293n00 & 22 55 26.1 & $+$29 05 28 & 4.258 & 5.416 & 5.68 &    0.74 & 1.12 & R & \nodata & \nodata & \nodata & \phn \\ 
WARPJ2302.7$+$0845 & rp700423n00 & 23 02 46.4 & $+$08 45 20 & 3.803 & 5.225 & 5.28 &    0.71 & 0.19 & R & \nodata & \nodata & \nodata & \phn \\ 
WARPJ2302.8$+$0843 & rp700423n00 & 23 02 48.1 & $+$08 43 50 & 4.628 & 5.239 & 4.86 &    0.72 & 0.92 & R & Y & CLG & 0.722 & \phn \\ 
WARPJ2302.9$+$0839 & rp700423n00 & 23 02 54.7 & $+$08 39 04 & 4.470 & 5.571 & 4.86 &    0.75 & 0.22 & R & \nodata & \nodata & \nodata & \phn \\ 
WARPJ2303.0$+$0846 & rp700423n00 & 23 03 01.0 & $+$08 46 55 & 5.453 & 6.037 & 4.86 &    0.81 & 0.71 & \nodata & \nodata & \nodata & \nodata & \phn \\ 
WARPJ2304.0$-$0837 & rp701250n00 & 23 04 00.5 & $-$08 37 52 & 5.346 & 6.553 & 3.51 &    0.86 & 0.01 & \nodata & \nodata & \nodata & \nodata & \phn \\ 
WARPJ2319.5$+$1226 & rp200474n00 & 23 19 34.6 & $+$12 26 21 & 21.225 & 31.721 & 4.42 &    4.15 & 1.42 & R & Y & CLG & 0.124 & VMF~217,RIXOS \\ 
WARPJ2320.7$+$1659 & rp600439n00 & 23 20 46.1 & $+$16 59 45 & 3.430 & 5.279 & 4.38 &    0.70 & 1.44 & R & Y & BLEND & 0.499 & AGN~(z=1.8)+CLG~(z=0.50) \\ 
WARPJ2320.9$+$1715 & rp600439n00 & 23 20 57.8 & $+$17 15 10 & 14.326 & 15.559 & 4.68 &    2.47 & 0.79 & \nodata & \nodata & star & A9/F0 & HD~220091 \\ 
\enddata
\end{deluxetable}

\begin{deluxetable}{llccrrrrrccccll}
\rotate
\tablewidth{0pt}
\tabletypesize{\scriptsize}
\tablecaption{Other Sources\label{tab:lowflux}}
\tablehead{\colhead{Name} & \colhead{ROSAT} & \colhead{ $\alpha$ } & \colhead{ $\delta$ } & \colhead{ cr$_{raw}$ } & \colhead{ cr$_{corr}$ } & \colhead{n$_{H}$ } & \colhead{Flux} & \colhead{Extent} & \colhead{Im} & \colhead{Sp} & \colhead{Id} & \colhead{Redshift/ } & \colhead{Other Name/ }\\
\colhead{ } & \colhead{Field} & \colhead{(J2000)} & \colhead{(J2000)} & \colhead{ [10$^{-3}$] } & \colhead{ [10$^{-3}$] } & \colhead{ [10$^{20}$] } & \colhead{ [10$^{-13}$] } & \colhead{ } & \colhead{ } & \colhead{ } & \colhead{ } & \colhead{Sp. Type} & \colhead{Notes}\\
\colhead{(1)} & \colhead{(2)} & \colhead{(3)} & \colhead{(4)} & \colhead{(5)} & \colhead{(6)} & \colhead{(7)} & \colhead{(8)} & \colhead{(9)} & \colhead{(10)} & \colhead{(11)} & \colhead{(12)} & \colhead{(13)} & \colhead{(14)}}
\startdata
WARPJ0019.5$+$2156 & rp400322n00 & 00 19 34.7 & $+$21 56 24 & 3.261 & 4.091 & 4.28 &    0.54 & 1.16 & R & \nodata & \nodata & \nodata & \phn \\ 
WARPJ0038.5$+$3058 & rp201045n00 & 00 38 34.2 & $+$30 58 22 & 3.240 & 3.794 & 5.60 &    0.52 & 0.77 & \nodata & Y & QSO & 1.326 & \phn \\ 
WARPJ0045.8$+$0107 & rp700377n00 & 00 45 50.3 & $+$01 07 24 & 3.447 & 4.353 & 2.62 &    0.56 & 0.94 & R & \nodata & \nodata & \nodata & \phn \\ 
WARPJ0110.7$-$3813 & rp700794n00 & 01 10 44.9 & $-$38 13 28 & 3.538 & 4.477 & 1.78 &    0.56 & 1.16 & R & \nodata & \nodata & \nodata & \phn \\ 
WARPJ0207.1$+$1504 & rp300003n00 & 02 07 06.8 & $+$15 04 54 & 3.233 & 4.303 & 6.18 &    0.60 & 0.72 & R & \nodata & \nodata & \nodata & RIXOS \\ 
WARPJ0216.9$-$1753 & rp900352n00 & 02 16 57.9 & $-$17 53 58 & 3.169 & 3.838 & 2.98 &    0.49 & 1.12 & R & \nodata & \nodata & \nodata & \phn \\ 
WARPJ0234.2$-$0356 & rp700350n00 & 02 34 13.3 & $-$03 56 51 & 3.624 & 5.114 & 2.59 &    0.64 & 1.31 & R & Y & CLG & 0.447 & \phn \\ 
WARPJ0236.0$-$5224 & rp701356n00 & 02 36 04.2 & $-$52 24 50 & 3.472 & 4.064 & 3.08 &    0.53 & 1.09 & R & \nodata & \nodata & \nodata & VMF~027 \\ 
WARPJ0236.7$-$5209 & rp701356n00 & 02 36 47.9 & $-$52 09 22 & 3.584 & 4.007 & 3.07 &    0.52 & 0.14 & R & \nodata & \nodata & \nodata & \phn \\ 
WARPJ0255.3$+$0004 & rp701403n00 & 02 55 23.1 & $+$00 04 38 & 3.627 & 4.584 & 6.48 &    0.64 & 1.18 & R & \nodata & \nodata & \nodata & \phn \\ 
WARPJ0255.4$-$0002 & rp701403n00 & 02 55 27.9 & $-$00 02 24 & 3.190 & 4.581 & 6.48 &    0.64 & 1.35 & R & \nodata & blend & \nodata & AGN~(Q~0252-0014)+? \\ 
WARPJ0347.8$-$1158 & rp900492n00 & 03 47 50.3 & $-$11 58 43 & 3.002 & 3.911 & 3.80 &    0.51 & 1.04 & R & \nodata & \nodata & \nodata & \phn \\ 
WARPJ0412.4$-$1724 & rp900246n00 & 04 12 30.0 & $-$17 24 38 & 3.818 & 4.916 & 2.52 &    0.63 & 1.18 & R & \nodata & \nodata & \nodata & \phn \\ 
WARPJ0414.4$-$1236 & rp900242n00 & 04 14 25.2 & $-$12 36 41 & 3.970 & 4.614 & 3.70 &    0.61 & 0.03 & R & \nodata & \nodata & \nodata & \phn \\ 
WARPJ0414.7$-$1240 & rp900242n00 & 04 14 43.2 & $-$12 40 04 & 3.008 & 3.714 & 3.70 &    0.49 & 1.08 & R & \nodata & \nodata & \nodata & \phn \\ 
WARPJ0440.6$-$1636 & rp900017n00 & 04 40 39.1 & $-$16 36 05 & 3.147 & 4.228 & 4.69 &    0.57 & 1.24 & R & \nodata & \nodata & \nodata & \phn \\ 
WARPJ1311.1$+$3711 & rp700796n00 & 13 11 11.0 & $+$37 11 57 & 3.460 & 3.988 & 1.11 &    0.49 & 1.16 & R & \nodata & \nodata & \nodata & \phn \\ 
WARPJ1318.8$+$4457 & rp900325n00 & 13 18 52.8 & $+$44 57 42 & 3.734 & 4.262 & 1.54 &    0.53 & 1.03 & R & \nodata & \nodata & \nodata & \phn \\ 
WARPJ1349.0$+$0750 & rp300028n00 & 13 49 02.3 & $+$07 50 44 & 3.169 & 3.852 & 2.05 &    0.49 & 1.06 & R & \nodata & \nodata & \nodata & \phn \\ 
WARPJ1407.6$+$3415 & rp700117n00 & 14 07 40.5 & $+$34 15 18 & 3.404 & 4.729 & 1.20 &    0.58 & 1.30 & R & Y & CLG & 0.577 & \phn \\ 
WARPJ1407.7$+$2824 & rp700061n00 & 14 07 45.6 & $+$28 24 42 & 3.057 & 4.322 & 1.46 &    0.54 & 0.92 & R & \nodata & qso & 1.127 & CRSS~J1407.7+2824 \\ 
WARPJ1415.0$+$1119 & rp700122n00 & 14 15 04.5 & $+$11 19 35 & 2.671 & 3.649 & 1.80 &    0.46 & 1.02 & R & \nodata & qso & 0.538 & CRSS~J1415.0+1119 \\ 
WARPJ1415.6$+$3622 & rp600462n00 & 14 15 39.0 & $+$36 22 11 & 4.010 & 5.098 & 0.99 &    0.63 & 1.19 & R & \nodata & gal & 0.0159 & KUG~1413+366 \\ 
WARPJ1501.0$-$0824 & rp200510n00 & 15 01 04.3 & $-$08 24 56 &  3.029 &  3.419 & 7.33 & 0.49 & 0.69    & R & N & \nodata & 0.51  & estimated redshift     \\ 
WARPJ1515.6$+$4346 & rp200965n00 & 15 15 31.9 & $+$43 46 18 & 15.821 & 24.478 & 1.96 & 2.98 & 1.48    & R & Y & CLG     & 0.135 & VMF~169  \\ 
WARPJ1517.9$+$3127 & rp201018n00 & 15 17 55.9 & $+$31 27 35 &  3.016 &  4.570 & 1.90 & 0.57 & 1.43    & R & Y & CLG     & 0.744 & \nodata  \\ 
WARPJ1535.1$+$2336 & rp701411n00 & 15 35 06.0 & $+$23 36 55 & 4.190 & 4.589 & 4.29 &    0.61 & 1.19 & \nodata & Y & AGN & 1.25 & \phn \\ 
WARPJ1537.7$+$1201 & rp400117n00 & 15 37 43.0 & $+$12 01 14 &  5.512 & 10.520 & 3.44 & 1.29 & \nodata & R & Y & CLG     & 0.136 & VMF~171  \\ 
WARPJ1552.1$+$2017 & rp701213n00 & 15 52 07.8 & $+$20 17 39 & 3.196 & 3.722 & 3.70 &    0.49 & 1.00 & R & \nodata & \nodata & \nodata & \phn \\ 
WARPJ1605.2$+$2541 & rp300021n00 & 16 05 17.2 & $+$25 41 23 & 3.306 & 4.405 & 4.62 &    0.59 & 0.21 & R & \nodata & qso & 1.071 & CRSS~J1605.2+2541 \\ 
WARPJ1605.6$+$2543 & rp300021n00 & 16 05 39.7 & $+$25 43 08 & 3.380 & 4.287 & 4.62 &    0.57 & 1.15 & R & \nodata & gal & 0.278 & CRSS~J1605.6+2543 \\ 
WARPJ1717.4$+$4319 & rp300180n00 & 17 17 29.8 & $+$43 19 41 & 3.662 & 4.446 & 2.19 &    0.56 & 0.40 & \nodata & Y & AGN & 0.632 & \phn \\ 
WARPJ2108.6$-$0507 & rp300389n00 & 21 08 39.9 & $-$05 07 27 & 2.213 & 3.132 & 5.57 &    0.41 & 1.29 & R & Y & CLG & 0.222 & \phn \\ 
WARPJ2157.6$-$1506 & rp701420n00 & 21 57 40.2 & $-$15 06 58 & 3.674 & 4.650 & 3.75 &    0.61 & 1.16 & R & \nodata & \nodata & \nodata & \phn \\ 
WARPJ2224.5$-$0208 & rp701018n00 & 22 24 31.7 & $-$02 08 05 & 3.011 & 3.671 & 4.95 &    0.50 & 0.91 & BR & \nodata & \nodata & \nodata & \phn \\ 
WARPJ2224.6$-$0218 & rp701018n00 & 22 24 41.2 & $-$02 18 26 & 3.110 & 4.040 & 4.95 &    0.55 & 1.12 & R & \nodata & \nodata & \nodata & \phn \\ 
WARPJ2239.5$-$0600 & rp701206n00 & 22 39 35.4 & $-$06 00 17 & 3.330 & 5.155 & 3.96 &    0.63 & 1.46 & BR & Y & CLG & 0.174 & VMF~209 \\ 
WARPJ2253.9$+$2911 & rp400293n00 & 22 53 54.5 & $+$29 11 46 & 3.126 & 3.603 & 5.68 &    0.50 & 1.86 & \nodata & Y & \nodata & \nodata & \phn \\ 
WARPJ2315.2$-$0529 & rp300220n00 & 23 15 14.6 & $-$05 29 47 & 3.647 & 4.239 & 3.59 &    0.55 & 1.08 & R & \nodata & \nodata & \nodata & \phn \\ 
WARPJ2316.4$-$0526 & rp300220n00 & 23 16 25.1 & $-$05 26 12 & 3.673 & 4.613 & 3.59 &    0.60 & 1.16 & R & \nodata & \nodata & \nodata & \phn \\ 
WARPJ2318.0$+$1238 & rp200474n00 & 23 18 06.0 & $+$12 38 12 & 3.574 & 4.233 & 4.24 &    0.56 & 1.06 & R & \nodata & agn & 0.713 & RIXOS~F294~001,RIXOS \\ 
WARPJ2343.4$-$1443 & rp701205n00 & 23 43 27.6 & $-$14 43 02 & 3.369 & 4.906 & 2.20 &    0.62 & 1.36 & R & \nodata & \nodata & \nodata & \phn \\ 
WARPJ2343.4$-$1500 & rp701205n00 & 23 43 26.3 & $-$15 00 51 & 3.181 & 4.173 & 2.20 &    0.53 & 1.22 & R & \nodata & \nodata & \nodata & \phn \\ 
\enddata
\end{deluxetable}

\begin{deluxetable}{lccrrrrrlcll}
\rotate
\tablewidth{0pt}
\tabletypesize{\scriptsize}
\tablecaption{WARPS-I Statistically Complete, Flux Limited Sample\label{tab:cluster}}
\tablehead{\colhead{Name} & \colhead{ $\alpha$ } & \colhead{ $\delta$ } & \colhead{Extent} & \colhead{n$_{H}$ } & \colhead{Flux} & \colhead{ log L$_{\rm x}$ } & \colhead{ r$_{\rm c}$ } & \colhead{z} & \colhead{ n$_{z}$ } & \colhead{Flags} & \colhead{Other Name/ }\\
\colhead{ } & \colhead{(J2000)} & \colhead{(J2000)} & \colhead{ } & \colhead{ [10$^{20}$] } & \colhead{ [10$^{-13}$] } & \colhead{  } & \colhead{ [$\arcsec$] } & \colhead{ } & \colhead{ } & \colhead{ } & \colhead{Notes}\\
\colhead{(1)} & \colhead{(2)} & \colhead{(3)} & \colhead{(4)} & \colhead{(5)} & \colhead{(6)} & \colhead{(7)} & \colhead{(8)} & \colhead{(9)} & \colhead{(10)} & \colhead{(11)} & \colhead{(12)}}
\startdata
WARPJ0111.6$-$3811 & 01 11 36.0 & $-$38 11 08 & 1.63 & 1.78 &    0.83 &   42.75 &    18.4 & 0.121 & 3 & \phn  & VMF~009 \\ 
WARPJ0144.5$+$0212 & 01 44 30.3 & $+$02 12 24 & 1.75 & 2.91 &    1.76 &   43.34 &    42.5 & 0.165 & 2 & \phn  & VMF~019 \\ 
WARPJ0206.3$+$1511 & 02 06 23.7 & $+$15 11 03 & 1.39 & 6.18 &    0.86 &   43.40 &    19.2 & 0.251 & 2 & \phn  & VMF~022,RIXOS \\ 
WARPJ0210.4$-$3929 & 02 10 26.7 & $-$39 29 27 & 1.76 & 1.43 &    0.90 &   43.49 &    39.4 & 0.273: & 2 & \phn  & VMF~025 \\ 
WARPJ0216.5$-$1747 & 02 16 32.8 & $-$17 47 11 & 1.60 & 2.98 &    1.75 &   44.41 &    39.1 & 0.578 & 3 & \phn c & \phn \\ 
WARPJ0228.1$-$1005 & 02 28 11.7 & $-$10 05 30 & 1.75 & 2.51 &    3.23 &   43.50 &    42.5 & 0.149 & 2 & \phn c & VMF~026 \\ 
WARPJ0238.0$-$5224 & 02 38 01.2 & $-$52 24 41 & 1.28 & 3.07 &    5.18 &   43.62 &    24.8 & 0.135 & 1 & \phn  & A3038, VMF~028, SHARC \\ 
WARPJ0250.0$+$1908 & 02 50 03.5 & $+$19 08 00 & 1.56 & 9.87 &    1.68 &   43.05 &    21.2 & 0.122 & 2 & \phn  & SHARC \\ 
WARPJ0348.1$-$1201 & 03 48 08.7 & $-$12 01 34 & 1.38 & 3.80 &    0.97 &   44.02 &    15.3 & 0.488 & 5 & \phn  & \phn \\ 
WARPJ0951.7$-$0128 & 09 51 47.0 & $-$01 28 30 & 1.73 & 3.84 &    0.73 &   44.03 &    24.2 & 0.568 & 1 & \phn c & VMF~076 \\ 
WARPJ1113.0$-$2615 & 11 13 05.7 & $-$26 15 36 & 1.21 & 5.47 &    0.75 &   44.26 &     6.7 & 0.725 & 2 & \phn  & \phn \\ 
WARPJ1406.2$+$2830 & 14 06 16.1 & $+$28 30 55 & 1.39 & 1.46 &    0.87 &   44.07 &    11.8 & 0.546 & \nodata & \phn  & VMF~153 \\ 
WARPJ1406.9$+$2834 & 14 06 55.4 & $+$28 34 25 & 1.58 & 1.46 &    2.25 &   43.14 &    22.7 & 0.117 & 3 & \phn c & VMF~154, SHARC \\ 
WARPJ1415.1$+$3612 & 14 15 11.1 & $+$36 12 03 & 1.36 & 1.14 &    1.13 &   44.39 &    11.6 & [0.7] & \nodata & \phn  & FIRST \\ 
WARPJ1418.5$+$2511 & 14 18 31.8 & $+$25 10 59 & 1.18 & 1.93 &    6.20 &   44.37 &    11.9 & 0.296 & 3 & \phn  & VMF~159, SHARC \\ 
WARPJ1552.2$+$2013 & 15 52 12.5 & $+$20 13 32 & 2.04 & 3.70 &    3.29 &   43.43 &    35.7 & 0.135 & 3 & \phn  & VMF~175, SHARC \\ 
WARPJ2038.4$-$0125 & 20 38 29.6 & $-$01 25 11 & 1.71 & 7.02 &    0.70 &   44.17 &    20.3 & 0.679 & 2 & \phn  & \phn \\ 
WARPJ2108.8$-$0516 & 21 08 51.0 & $-$05 16 32 & 1.40 & 5.57 &    1.89 &   43.93 &    29.7 & 0.317 & 10 & \phn  & VMF~200 \\ 
WARPJ2239.4$-$0547 & 22 39 24.8 & $-$05 47 10 & 1.30 & 4.01 &    2.88 &   43.87 &    16.3 & 0.242 & 6 & \phn  & A2465S, VMF~208 \\ 
WARPJ2239.6$-$0543 & 22 39 39.6 & $-$05 43 15 & 1.65 & 4.01 &    4.04 &   44.02 &    43.7 & 0.243 & 6 & \phn  & A2465N, VMF~210 \\ 
WARPJ2302.8$+$0843 & 23 02 48.1 & $+$08 43 50 & 0.92 & 4.86 &    0.72 &   44.23 &     0.1 & 0.722 & 1 & \phn c & \phn \\ 
WARPJ2319.5$+$1226 & 23 19 34.6 & $+$12 26 21 & 1.42 & 4.42 &    4.15 &   43.45 &    28.8 & 0.124 & 3 & \phn  & VMF~217,RIXOS  
\enddata
\end{deluxetable}

\begin{deluxetable}{lccrrrrrlcll}
\rotate
\tablewidth{0pt}
\tabletypesize{\scriptsize}
\tablecaption{WARPS-I Additional Clusters and Candidates\label{tab:other}}
\tablehead{\colhead{Name} & \colhead{ $\alpha$ } & \colhead{ $\delta$ } & \colhead{Extent} & \colhead{n$_{H}$ } & \colhead{Flux} & \colhead{ log L$_{\rm x}$ } & \colhead{ r$_{\rm c}$ } & \colhead{z} & \colhead{ n$_{z}$ } & \colhead{Flags} & \colhead{Other Name/ }\\
\colhead{ } & \colhead{(J2000)} & \colhead{(J2000)} & \colhead{ } & \colhead{ [10$^{20}$] } & \colhead{ [10$^{-13}$] } & \colhead{  } & \colhead{ [$\arcsec$] } & \colhead{ } & \colhead{ } & \colhead{ } & \colhead{Notes}\\
\colhead{(1)} & \colhead{(2)} & \colhead{(3)} & \colhead{(4)} & \colhead{(5)} & \colhead{(6)} & \colhead{(7)} & \colhead{(8)} & \colhead{(9)} & \colhead{(10)} & \colhead{(11)} & \colhead{(12)}}
\startdata
WARPJ0022.0$+$0422 & 00 22 03.4 & $+$04 22 38 & 1.81 & 2.87 &    0.72 &   43.74 &    28.8 & 0.407 & 5 & 8c & GHO~00190.5+0405 \\ 
WARPJ0023.1$+$0421 & 00 23 06.0 & $+$04 21 13 & 1.41 & 2.87 &    0.76 &   43.86 &    15.6 & 0.453 & 3 & 8 & \phn \\ 
WARPJ0210.2$-$3932 & 02 10 13.8 & $-$39 32 50 & 1.77 & 1.43 &    0.46 &   42.89 &    19.8 & [0.190] & \nodata &  $<$  & VMF~024, estimated~redshift \\ 
WARPJ0234.2$-$0356 & 02 34 13.3 & $-$03 56 51 & 1.31 & 2.59 &    0.64 &   43.78 &    12.4 & 0.447 & 5 &  $<$  & \phn \\ 
WARPJ0236.0$-$5224 & 02 36 04.2 & $-$52 24 50 & 1.09 & 3.08 &    0.53 & \nodata &     4.2 & \nodata & - & ? & VMF~027 \\ 
WARPJ0255.3$+$0004 & 02 55 23.1 & $+$00 04 38 & 1.18 & 6.48 &    0.64 & \nodata &     5.3 & \nodata & - & ? & \phn \\ 
WARPJ1407.6$+$3415 & 14 07 40.5 & $+$34 15 18 & 1.30 & 1.20 &    0.58 & \nodata &    11.8 & 0.577 & 1 & ? & \phn \\ 
WARPJ1415.3$+$2308 & 14 15 18.4 & $+$23 08 57 & 1.50 & 1.79 &    0.57 &   42.02 &    15.4 & 0.064 & 5 & 8 & \phn \\ 
WARPJ1416.4$+$2315 & 14 16 26.9 & $+$23 15 32 & 1.32 & 1.85 &   12.79 &   44.02 &    46.1 & 0.137 & 2 & 8c & SHARC \\ 
WARPJ1501.0$-$0824 & 15 37 43.0 & $+$12 01 14 & 0.69    & 7.33  & 0.49    & 43.79   & \nodata & [0.51]  & - & ?    & estimated redshift    \\ 
WARPJ1515.5$+$4346 & 15 15 31.9 & $+$43 46 18 & \nodata & 1.96  & \nodata & \nodata & \nodata & 0.135 & 3 & b    & VMF~169  \\ 
WARPJ1517.9$+$3127 & 15 17 55.9 & $+$31 27 35 & 1.43    & 1.90  & 0.57    & 44.16   & 0.41    & 0.744 & 2 & $<$  & \phn  \\ 
WARPJ1537.7$+$1201 & 15 37 43.0 & $+$12 01 14 & \nodata & 3.44  & 1.20    & 43.01   & \nodata & 0.134 & 3 & b    & VMF~171  \\ 
WARPJ2108.6$-$0507 & 21 08 39.9 & $-$05 07 27 & 1.29 & 5.57 &    0.41 &   42.98 &    10.5 & 0.222 & 3 &  $<$  & \phn \\ 
WARPJ2146.0$+$0423 & 21 46 05.5 & $+$04 23 13 & 1.33 & 5.42 &    1.84 &   44.36 &    20.0 & 0.532 & 4 & 8 & VMF~204, GHO~2143+0408 \\ 
WARPJ2239.5$-$0600 & 22 39 35.4 & $-$06 00 17 & 1.46 & 3.96 &    0.63 &   42.95 &    20.2 & 0.174 & 2 &  $<$  & VMF~209 \\ 
WARPJ2320.7$+$1659 & 23 20 46.1 & $+$16 59 45 & 1.44 & 4.38 &    0.70 & \nodata &    18.5 & 0.499 & 3 & $<$c & AGN~(z=1.8)+CLG~(z=0.50) \\ 
\enddata
\end{deluxetable}

\begin{deluxetable}{lccrrrl}
\tablewidth{0pt}
\tabletypesize{\scriptsize}
\tablecaption{Radio Sources within 2$\arcmin$ of WARPS Cluster Candidates\label{tab:radio}}
\tablehead{\colhead{Name} & \colhead{ $\alpha$ } & \colhead{ $\delta$ } & \colhead{Error} & \colhead{Dist} & \colhead{Flux} & \colhead{Comments}\\
\colhead{ } & \colhead{(J2000)} & \colhead{(J2000)} & \colhead{ [$\arcsec$] } & \colhead{ [$\arcsec$] } & \colhead{ [mJy] } & \colhead{ }}
\startdata
WARPJ0144.5$+$0212:R1 & 01 44 29.89 & $+$02 12 43.3 & 1 & 20 & 80.6 & Just north of BCG \\ 
WARPJ0144.5$+$0212:R2 & 01 44 33.08 & $+$02 11 41.6 & 4 & 59 & 8.9 & \phn \\ 
WARPJ0206.3$+$1511:R1 & 02 06 20.44 & $+$15 10 56.1 & 1 & 48 & 50.4 & \phn \\ 
WARPJ0216.5$-$1747:R1 & 02 16 32.22 & $-$17 45 17.2 & 3 & 115 & 7.2 & \phn \\ 
WARPJ0228.1$-$1005:R1 & 02 28 11.93 & $-$10 05 32.7 & 4 & 5 & 7.0 & BCG \\ 
WARPJ0348.1$-$1201:R1 & 03 48 09.29 & $-$12 01 59.0 & 1 & 26 & 39.9 & cluster member \\ 
WARPJ0951.7$-$0128:R1 & 09 51 46.39 & $-$01 28 32.8 & 4 & 10 & 6.2 & cluster member \& star in error circle \\ 
WARPJ1407.6$+$3415:R1 & 14 07 45.46 & $+$34 15 27.2 & 4 & 62 & 0.27 & may be a sidelobe of another source \\ 
WARPJ1415.1$+$3612:R1 & 14 15 10.87 & $+$36 12 07.0 & 6 & 5 & 5.4 & \phn \\ 
WARPJ1415.3$+$2308:R1 & 14 15 18.39 & $+$23 08 39.6 & 4 & 18 & 8.7 & \phn \\ 
WARPJ1416.4$+$2315:R1 & 14 16 27.79 & $+$23 15 35.3 & 10 & 12 & 4.1 & BCG? \\ 
WARPJ1418.5$+$2511:R1 & 14 18 32.57 & $+$25 12 03.9 & 1 & 65 & 38.9 & \phn \\ 
WARPJ1501.0$-$0824:R1 & 15 01 03.22 & $-$08 24 39.9 & 6 & 23 & 9.6 & \phn \\ 
WARPJ2038.4$-$0125:R1 & 20 38 24.76 & $-$01 24 02.1 & 21 & 100 & 7.9 & Very extended \\ 
WARPJ2038.4$-$0125:R2 & 20 38 35.49 & $-$01 25 58.4 & 1 & 100 & 31.1 & \phn \\ 
WARPJ2108.6$-$0507:R1 & 21 08 38.81 & $-$05 07 43.4 & 9 & 22 & 3.7 & \phn \\ 
WARPJ2108.8$-$0516:R1 & 21 08 48.87 & $-$05 16 05.1 & 3 & 41 & 9.6 & \phn \\ 
WARPJ2146.0$+$0423:R1 & 21 46 05.92 & $+$04 22 44.1 & 11 & 30 & 3.6 & \phn \\ 
WARPJ2239.6$-$0543:R1 & 22 39 40.32 & $-$05 43 23.0 & 6 & 14 & 8.4 & BCG \\ 
WARPJ2302.8$+$0843:R1 & 23 02 48.10 & $+$08 43 50.8 & 1 & 0 & 25.6 & \phn \\ 
WARPJ2320.7$+$1659:R1 & 23 20 44.42 & $+$17 00 18.9 & 10 & 41 & 2.8 & background AGN? \\ 
\enddata
\end{deluxetable}

\label{tab:radio}

\end{document}